\documentclass[a4paper,preprint,superscriptaddress,amsfonts, amssymb, amsmath, showkeys]{revtex4-1}
\usepackage[english]{babel}
\usepackage{natbib}
\usepackage[utf8]{inputenc}
\usepackage{graphicx}
\usepackage{tikz-cd}
\usepackage{siunitx}
\usepackage{amsthm}
\usepackage{mathtools}
\usepackage{physics}
\usepackage{xcolor}
\usepackage{graphicx}
\usepackage[left=23mm,right=13mm,top=35mm,columnsep=15pt]{geometry} 
\usepackage{adjustbox}
\usepackage{placeins}
\usepackage[T1]{fontenc}
\usepackage{lipsum}
\usepackage{csquotes}
\usepackage{graphicx}
\usepackage{float}
\usepackage[font=small]{caption}
\usepackage{subcaption}
\usepackage{blindtext}
\usepackage{amsthm}
\usepackage{enumerate}
\usepackage{framed}
\usepackage{amssymb}
\usepackage[makeroom]{cancel}
\usepackage{amsmath,mathrsfs}
\usepackage{wasysym}
\usepackage{hhline}
\usepackage{ulem}
\usepackage{simplewick}
\usepackage{feynmp}
\usepackage{hyperref}
\usepackage{tikz}
\everymath{\displaystyle}

\everymath{\displaystyle}

\newcommand{\Hess}{\mathrm{Hess}}
\newcommand{\bi}{\begin{itemize}}
\newcommand{\ei}{\end{itemize}}
\newcommand{\D}{\mathrm{d}}
\newcommand{\bv}{\bar{v}}
\newcommand{\Christoffel}[3]{\Gamma^{#1}_{#2 #3}}
\newcommand{\ChristoffelTilde}[3]{\widetilde{\Gamma}^{#1}_{#2 #3}}
\newcommand{\LevelSetfunc}[2]{\Sigma_{#1}^{#2}}
\newcommand{\Mfunc}[2]{M_{#1}^{#2}}

\newcommand{\intSigmaV}[2]{\int_{\LevelSetfunc{#1}{#2}}}

\newcommand{\R}{\mathbb R}

\newcommand{\T}{\mathcal{T}_f}

\newtheorem{theorem}{Theorem}[section]

\newtheorem{corollary}[theorem]{Corollary}
\newtheorem{definition}{Definition}[section]

\newtheorem{remark}{Remark}

\begin{document}
\title{Configurational microcanonical statistical mechanics from Riemannian geometry of equipotenital level sets}

\author{Matteo Gori}
\email[Correspondence email address: ]{matteo.gori@uni.lu}
\homepage[Current Address: ]{Department of Physics and Materials Science, University of Luxembourg, L-1511 Luxembourg City, Luxembourg}
\affiliation{Department of Physics and Materials Science, University of Luxembourg, L-1511 Luxembourg City, Luxembourg}
\affiliation{Aix Marseille Univ, Universit\'{e} de Toulon, CNRS, CPT, Marseille, France}
\affiliation{CNRS Centre de Physique Th\'{e}orique UMR7332, 13288 Marseille, France}
\date{\today} 

\begin{abstract}
In the present work, we present a detailed discussion of a Riemannian metric structure originally introduced in [Gori et al., \textit{J. Stat. Mech.}, \textbf{9} 093204 (2018)] on the configuration space and on phase space allowing to interpret the derivatives of the 
configurational microcanonical entropy and of the canonical
entropy in terms of integrals of extrinsic 
geoemtrical quantities associated to the
equipotential level sets.
\end{abstract}

\maketitle

\section{Introduction}
In the last decades, a field of research has been 
established aiming at investigating how thermodynamic 
properties of classical systems arise from
the geometrical properties of the equipotential level sets in configuration space\cite{pettini2007geometry,franzosi2011microcanonical}. For instance, the  so-called Topological Theory (TT) on the
origin of phase transitions (PT) in classical systems,
developed in the last three decades, represents one of the 
conceptual frameworks that have contributed to 
orienting the research in this field and have provided 
a generalization of the statistical mechanical description
of phase transitions in small or mesoscopic systems.
According to TT, the singularities of  thermodynamic 
potentials - arising in the thermodynamic limit in the 
canonical and grancanonical ensembles - would be induced by
suitable topological changes of some submanifolds of 
configuration space.
These same topological changes can occur for any finite number of degrees of freedom. This theory, supported by 
many pieces of evidence ranging from numerical simulations to exact 
analytic computations carried on different statistical 
models have been rigorously rooted in two 
theorems that associate topological changes of the 
equipotential level sets of configuration space with the 
loss of analyticity of microcanonical configuration 
entropy. 
In particular, one of these theorems named \textit{Necessity 
theorem}, states that in its original formulation that if 
all the equipotential level sets in a certain interval of 
specific potential energy $[\bar{v}_1, \bar{v}_2]$ are 
diffeomorphic among them at any finite, then the system 
cannot undergo any phase transition in the corresponding 
interval of temperatures for short-range regular potential.
According to Morse Theory, this statement can be rephrased
as follows: the absence of critical points of the potential
energy in the specific potential energy $[\bar{v}_1, \bar{v}_2]$ at any finite N implies the absence of phase 
transition in the corresponding interval of temperatures.
In the last decade, a counterexample to the original
formulation of the \textit{Necessity Theorem} has been 
found: for the $\phi^4$ model on a 2D lattice and nearest 
neighbors interactions, there are no critical points of
the potential energy for the critical value of the specific potential energy $\bar{v}_c$ corresponding to
the second-order phase transition observed at the critical
temperature $T_c$ in the thermodynamic limit.
These findings oriented the research towards a refinement of the 
topological theory leading to a new formulation of the 
\textit{Necessity Theorem} including the hypothesis of asymptotic
diffeomorphicity of equipotential level sets in thermodynamic 
limit. In the conceptualization process leading to such a
development of the Topological Theory (TT), it appears to be a key step
to search for a clear connection between the Riemannian geometric 
properties of the regular equipotential level sets and the 
derivatives of the  configurational microcanonical  entropy with 
respect to specific potential energy. The natural kinetic energy metric in configuration space does not provide a suited framework for this purpose.\\
In the present work, we present a method to define a Riemannian 
metric structure both in configuration space in the absence 
of critical points of potential energy allowing to provide clear 
identification of the derivatives of the configurational 
microcanonical entropy with integrals of extrinsic geometric 
curvature over equipotential  (isoenergetic) level sets.\\
\section{Geometry of regular potential energy level sets in configuration space}
\label{sec:geometry_regularlevelsets}

\subsection{Microcanonical configurational statistical mechanics from differential topology of regular equipotential level sets}

We consider in what follows the configurational microcanonical ensemble\footnote{The same considerations apply to the classical microcanonical ensemble
where the specific energy is fixed, simply replacing the configuration
space $\Lambda_{q}$ with the phase space $\Lambda_{p,q}$ and the specific
potential energy $\overline{V}_N$ (with fixed value $\bv$) with the
Hamiltonian representing the energy per degree of freedom
$\overline{\mathcal{H}}_N$ (with fixed value $\overline{\epsilon}$).}
$(\Lambda_{q},\rho_N(\mathbf{q};\overline{v}))$ where the constraint is obtained by fixing the value of some specific potential energy function $\overline{V}_{N}:\Lambda_q \longrightarrow\mathbb{R}$ and the corresponding microcanonical configurational density function 
is given by
\begin{equation}
\label{eq:GenMicrocan_sp}
\rho_N(\mathbf{q};\overline{v})=\dfrac{\delta\left(\overline{V}_N (\mathbf{q})-\overline{v}\right)}{\int_{\mathcal{X}^N}\delta\left(\overline{V}_N (\mathbf{q})-\overline{v}\right)\D\mathrm{Vol}_{g}}.
\end{equation}
where $g$ is a natural metric structure in configuration space\footnote{The introduction of a metric space is an arbitrary operation and not always the euclidean one is the best choice. For instance for a system with angular generalized coordinates $\theta_i\in[0;2\pi)$ the torus metric could be more appropriate.} and $\D\mathrm{Vol}_g$
the associated Riemannian volume form.
The normalization constant in \eqref{eq:GenMicrocan_sp} is the microcanonical partition function according to Boltzmann's definition:
\begin{equation}
\label{eq:MicroPartitionFunction}
\Omega_{N,\mathrm{Boltz}}(\overline{v})=\dfrac{\partial}{\partial\overline{v}}\Omega_{N,\mathrm{Gibbs}}(\overline{v})=\dfrac{\partial}{\partial\overline{v}}\int_{\Lambda_q}\Theta(\overline{V}_N(\mathbf{q})-\overline{v})\D\mathrm{Vol}_{g}=\int_{\Lambda_{q}}\delta\left(\overline{V}_N (\mathbf{q})-\overline{v}\right)\D\mathrm{Vol}_{g}
\end{equation}
where $\Theta(x)$ is the Heaviside step function.\\
In analogy with the usual definitions in statistical microcanonical ensemble, the configurational  microcanonical entropy density function is given by:
\begin{equation}
\label{eq:BoltzmannEntropy}
\overline{S}_{N,\mathrm{Boltz}}=\dfrac{1}{N}\ln\Omega_{N,\mathrm{Boltz}}(\overline{v})
\end{equation}

As a consequence of eq.\eqref{eq:BoltzmannEntropy}, the microcanonical volume of the level sets of $\overline{V}_N$ as a function of $\bv$ contains the whole information on the thermodynamics of the system.\\
In what follows, the thermodynamic properties of the configurational microcanonical ensemble are expressed as integrals of quantities associated with the vector field that generates the diffeomorphism among the level sets of $\overline{V}_N$.

\begin{definition}[Equipotential level sets]
We recall that the equipotential level sets $\LevelSetfunc{\bv}{\overline{V}_N}$ are defined as
\begin{equation}
\LevelSetfunc{\bv}{\overline{V}_N}=\left\{q\in\Lambda_{q}|\overline{V}_N(q)=\overline{v}\right\}
\end{equation}
\end{definition}

\begin{remark}[Compactness of level sets]
As $\overline{V}_N\in C^{\infty}(\Lambda_q)$ is a continuous function then
$\overline{V}_N^{-1}(\overline{v})\subseteq \Lambda_q$ is a compact set. 
\end{remark}

Let the gradient vector field $\boldsymbol{\mathrm{grad}}_{g}\overline{V}_N\in\mathfrak X(\Lambda_q)$ of the function $\overline{V}_N$ be
\begin{equation}
g(\boldsymbol{\mathrm{grad}}_{g}\overline{V}_N,\mathbf{X})=\D \overline{V}_N(X) \qquad \forall \mathbf{X}\in\mathfrak X(\Lambda_q) \,\, .
\end{equation}
\begin{remark}[Non critical level set]
If there are no critical points of $\overline{V}_N$ on the level set $\LevelSetfunc{\overline{v}}{\overline{V}_N}$, i.e.:

\begin{equation}
\label{eq:def_NonCriticalLevelSet}
\boldsymbol{\mathrm{grad}}_{g} \overline{V}_N \Bigr|_{\mathbf{q}}\neq\mathbf{0} \qquad
\forall q\in\Sigma_{\overline{v}}
\end{equation}
then $\LevelSetfunc{\overline{v}}{\overline{V}_N}$ is a regular compact hypersurface.
\end{remark}

Let consider a configuration space subset where there is no 
critical point, i.e.
\begin{equation}
\mathcal{B}_N=\left\{q\in\Lambda_q|\  \boldsymbol{\mathrm{grad}}_{g}\overline{V}
\Bigr|_{q}\neq\mathbf{0} \right\}.
\end{equation}
and  the set
\begin{equation}
\Mfunc{\left[\overline{v}_0,\overline{v}_1\right]}{\overline{V}_N}=\bigcup_{\bv\in\left[\bv_0,\bv_1\right]}\LevelSetfunc{\bv}{\overline{V}_N}
\end{equation}
and suppose that exist some $\overline{v}_0,\overline{v}_1$ such that $\Mfunc{\left[\bv_0,\bv_1\right]}{\overline{V}_N}\subset B$. This means that the one-form $\D\overline{V}_N$ is non-degenerate over $\Mfunc{\left[\bv_0,\bv_1\right]}{\overline{V}_N}$.\\
It follows from Froebenius' Theorem that a co-dimension one foliation can be defined
on $\Mfunc{\left[\bv_0,\bv_1\right]}{\overline{V}_N}$ through regular level sets of $\overline{V}_N$. Hence, it is possible to define a unit normal vector field to the 
equipotential hypersurfaces (the leaves of the foliation):

\begin{equation}
\boldsymbol{\nu}_N=\dfrac{\mathbf{grad}_g\overline{V}_N}{\|\mathbf{grad}_{g}\overline{V}_N\|_{g}}
\end{equation}

We stress that the absence of critical points of $\overline{V}_N$ over the manifold
$\Mfunc{\left[\overline{v}_0,\overline{v}_1\right]}{\overline{V}_N}$ has important consequences on the topology of the level sets therein: in particular, we will use the following well known result in differential topology:

\begin{theorem}[Regular interval theorem(\cite{hirsch1997differential}, p.153)]
\label{th:RegularIntervalTheorem}
Let $f:M\longrightarrow\left[a,b\right]$ be a $C^{r+1}$ map on a compact manifold $1\leq r \leq \omega$ (where $\omega$ means analytical). Suppose $f$ has no critical points and $f(\partial M)=\{a,b\}$. Then all level surfaces of $f$ are diffeomorphic.
\end{theorem}

In the proof of the same theorem in \cite{hirsch1997differential}, an explicit formulation is given for the vector field that generates the diffeomorphisms among the level sets and it is parametrized by the values taken by $f$ on them.\\
If the function $f$ is identified with $\overline{V}_N$ (as already mentioned in the previous Section) the vector field that generates the diffeomorphisms among the level sets, parametrized by $\bv$, is
\begin{equation}
\label{eq:def_HirschVector}
\overline{\boldsymbol{\xi}}_N=\dfrac{\boldsymbol{\mathrm{grad}}_{\mathbb{R}^N} \overline{V}_N}{\|\boldsymbol{\mathrm{grad}}_{g_{\mathbb{E}^N}} \overline{V}_N\|^2_{g_{\mathbb{E}^N}}}=\overline{\chi}_N \boldsymbol{\nu}_N \qquad \boldsymbol{\xi}_N\in \mathfrak{X}(M_{[v_0,v_1]}^N)\, .
\end{equation}
where we have introduced the symbol $\overline{\chi}_N$ for the norm of vector field $\overline{\boldsymbol{\xi}}_N$ in the ambient space ($\Lambda_q$, $g$), i.e.:
\begin{equation}
\label{eq:chi_def}
\overline{\chi}_N=\dfrac{1}{\|\boldsymbol{\mathrm{grad}}_g \overline{V}_N\|_g}=\|\overline{\boldsymbol{\xi}}_N\|_g\,\,.
\end{equation}
This means that the diffeomorphic flow generated by $\overline{\boldsymbol{\xi}}_N$ is normal to the level sets and is parametrized by the differences of $\overline{V}_N$ along the flow lines
\begin{equation}
\label{eq:def_HirschVector}
\mathrm{d}\overline{V}_N(\overline{\boldsymbol{\xi}}_N)=1 \, 
\end{equation}
More in explicit, this means that there exists a diffeomorphism flow
$\mathrm{Fl}:\LevelSetfunc{\bv_0}{\overline{V}_N}\times[0;\bv_1-\bv_0]\rightarrow \Mfunc{[\bv_0,\bv_1]}{\overline{V}_N}$ among the level
sets, s.t.
\begin{equation}
\begin{cases}
\mathrm{Fl}_t(p)\in\LevelSetfunc{\bv_0+t}{\overline{V}_N}\\
\dfrac{\D}{\D t}\left(f\circ\mathrm{Fl}_t\right)(p)\Biggr|_{t=(\bv-\bv_0)}=\left(\overline{\boldsymbol{\xi}}_N f\right)(\mathrm{Fl}_{(\bv-\bv_0)}(p)) \qquad \forall p\in\LevelSetfunc{\bv_0}{\overline{V}_N} \,\,\text{and}\,\,\forall \bv\in[\bv_0;\bv_1]\\
\end{cases}
\end{equation}
where $f$ is an arbitrary function of class $\mathit{C}^1$ defined over an open set of $\Lambda_q$ containing $\Mfunc{[\bv_0,\bv_1]}{\overline{V}_N}$.\\

In this differential topological framework it is possible
to express the microcanonical entropy and its derivatives in terms of integral of quantities related only to the vector field $\overline{\boldsymbol{\xi}}_N$: this establishes a link among the property of diffeomorphicity of the level sets in $\Mfunc{[\overline{v}_0,\overline{v}_1]}{\overline{V}_N}$ and the thermodynamic
behaviour of the system.
In particular the microcanonical partition function in eq.\eqref{eq:MicroPartitionFunction} can be rewritten in a more suitable form using the Coarea Formula \cite{federer2014geometric,Nic14_coarea} which generalizes Fubini's theorem. 

\begin{theorem}[Co-Area formula (\cite{Nic14_coarea} Corllary 1.4, p.5)]
\label{th:Coarea_Formula}
Suppose $\mathcal{M}$ is a $C^1$ manifold equipped with a $C^1$-metric $g$ and $f:M\longrightarrow\mathbb{R}$ is a function with no critical points. Then for any measurable function $\phi:\mathcal{M}\rightarrow\mathbb{R}$ we have
\begin{equation}
\int_{\mathcal{M}}\phi(p)\D\mathrm{Vol}_{g}=\int_{\mathbb{R}}\left(
\int_{\LevelSetfunc{t}{f}}\dfrac{\phi(p)}{\|\mathbf{grad}_{g}f\|_{g}}\D\sigma_{\LevelSetfunc{t}{f},g}\right)\D t
\end{equation}
where $\D\mathrm{Vol}_{g}$ is the Riemannian volume form on $\mathcal{M}$,
and $\D\sigma_{\LevelSetfunc{t}{f},g_{\mathcal{M}}}$ is its restriction over the 
regular level set $\LevelSetfunc{t}{f}$.
In particular, by setting $\phi=1$ it follows
\begin{equation}
\mathrm{Vol}_{g}(\mathcal{M})=\int_{\mathbb{R}}\left(\int_{\LevelSetfunc{t}{f}}\dfrac{\D\sigma_{\LevelSetfunc{t}{f},g}}{\|\mathbf{grad}_{g} f\|_{g}}\right)\D t
\end{equation} 
\end{theorem}

We can apply the Theorem \eqref{th:Coarea_Formula} to derive an useful expression for the configurational microcanonical partition function
$\Omega_{N}(\bv)$ for $\bv\in\mathcal{B}_N$.
In fact, let us consider two values  $\bv_0,\bv_1\in\mathcal{B}_N$
such that $\bv_0<\bv<\bv_1$, then the smooth function $\overline{V}_N$ 
has no critical points in $\Mfunc{\left[\overline{v}_0,\overline{v}_1\right]}{\overline{V}_N}$ and it follows

\begin{equation}
\begin{split}
\Omega_{N,\mathrm{Boltz}}(\bv)&=\dfrac{\partial}{\partial \bv'}\Omega_{N,\mathrm{Gibbs}}(\bv')\Biggr |_{\bv'=\bv}=\dfrac{\partial}{\partial \bv'}\left[\Omega_{N,\mathrm{Gibbs}}(\bv_0)+\int_{\Mfunc{[\bv_0,\bv']}{\overline{V}_N}}\D\mathrm{Vol}_{g}\right]\Biggr |_{\bv'=\bv}=\\
&=\dfrac{\partial}{\partial \bv'}\int_{\bv_{0}}^{\bv'}\left(\int_{\LevelSetfunc{t}{\overline{V}_N}}\dfrac{\D\sigma_{\LevelSetfunc{t}{\overline{V}_N},g}}{\|\mathbf{grad}_{g}\overline{V}_N\|_{g}}\right)\D t\Biggr |_{\bv'=\bv}=\int_{\LevelSetfunc{\bv}{\overline{V}_N}}\dfrac{\D\sigma_{\LevelSetfunc{\bv}{\overline{V}_N},g}}{\|\mathbf{grad}_{g}\overline{V}_N\|_{g}} \,.
\end{split}
\end{equation}

This very well known formula can be reinterpreted in order to make the vector field $\overline{\boldsymbol{\xi}}_N$ appear by simply using eq.\eqref{eq:chi_def}:
\begin{equation}
\label{eq: MicroCan_PartFunc_xi}
\Omega_{N,\mathrm{Boltz}}(\bv)=\int_{\LevelSetfunc{\bv}{\overline{V}_N}}\,\overline{\chi}_N\, \D\sigma_{\LevelSetfunc{\bv}{\overline{V}_N},g}=
\int_{\LevelSetfunc{\bv}{\overline{V}_N}}\D\mu^{N-1}_{\bv}
\end{equation}
where
\begin{equation}
\label{eq:def_microcanonical_areaform}
\D\mu^{N-1}_{\bv}=\overline{\chi}_N \D\sigma_{\LevelSetfunc{\bv}{\overline{V}_N},g}
\end{equation}
is the microcanonical area $(N-1)$-form for non critical energy level sets.
In what follows we refer only to the Boltzmann configurational microcanonical entropy, i.e. defined through the volume $\Omega_{N,\mathrm{Boltz}}(\bv)$.\\
As we have seen, the thermodynamic behaviour of a system is described by means of  
response functions which depend on the derivatives of the configurational microcanonical entropy and, consequently, of the configurational microcanonical partition function (volume).
So we need to calculate the derivatives of $\Omega_{N,\mathrm{Boltz}}(\bv)$ in eq.\eqref{eq: MicroCan_PartFunc_xi} with respect to the control parameter $\bv$ in the configurational microcanonical ensemble that we are considering, and then express these derivatives in terms of quantities directly related with the diffeomorphisms generating vector field.\\
The following result allows to do this
 

\begin{theorem}[Derivation of integral over regular level sets]
\label{Th:GeneralizedFederer}
Let $O$ an open bounded set of a Riemannian manifold $(\mathcal{M}^N,g)$ with a connection $\nabla$. Let $\psi\in \mathcal{C}^{p+1}(\overline{O})$ be constant on each connected component of the boundary $\partial\overline{O}$ and $f\in \mathcal{C}^{p}(O)$. Define $M_{]t,t'[}^N=\left\{x\in O |t<\psi(x)<t'\right\}$ and
\begin{equation}
F(v)=\intSigmaV{v}{\psi} \,f\,\D\sigma_{\LevelSetfunc{v}{\psi},g}
\end{equation}
where $\D\sigma_{\LevelSetfunc{v}{\psi},g}$ is the Riemannian area $N-1$-form induced over $\LevelSetfunc{v}{\psi}$. If $C>0$ exists such that for any $M_{]t,t'[}^{\psi}$, $\|\mathbf{grad}_{g}\psi(x)\|_{g}\geq C$ and the level sets $\LevelSetfunc{v}{\psi}$ of $\psi$ are without boundary, then for any $k$ such that $0\leq k\leq p $, for any $v\in]t,t'[$, one has
\begin{equation}
\dfrac{\D^k F}{\D v^k}(v)=\int_{\LevelSetfunc{v}{\psi}}\, A_{\psi,g}^{k}f\,\D\sigma_{\LevelSetfunc{v}{\psi},g}
\end{equation}
with
\begin{equation}
A_{\psi,g}f=\mathrm{div}_{g}\left(\boldsymbol{\nu}f\right)\dfrac{1}{\|\mathbf{grad}_{g}\psi(x)\|_{g}} \qquad \boldsymbol{\nu}=\dfrac{\mathbf{grad}_{g}\psi(x)}{\|\mathbf{grad}_{g}\psi(x)\|_{g}}
\end{equation}
\end{theorem}
\textit{proof}.

We prove the formula at the first order of derivation, namely for $k=1$
\begin{equation}
\label{eq:derivativefunctionLevelset}
\dfrac{\D F}{\D v}(v)=\dfrac{\D }{\D v}\intSigmaV{v}{\psi}\,A_{\psi,g}f\,\D\sigma_{\LevelSetfunc{v}{\psi},g}
\end{equation}
as the case for $k>1$ can be obtained by recursion.\\
The absence of critical points of $\psi$ implies that the level sets $\LevelSetfunc{v}{\psi}$ of $\psi$ determine a foliation of the open manifold  $M_{]t,t'[}^N$. Moreover all the level sets are diffeomorphic by after \autoref{th:RegularIntervalTheorem} and a
vector field ${\boldsymbol{\xi}}$ generating a family of one-parameter group of diffeomorphisms $\mathrm{Fl}_{t}$ parametrized by differences of values of $\psi$ can be found, i.e.
\begin{equation}
\boldsymbol{\xi}=\dfrac{\mathbf{grad}_{g}\psi(x)}{\|\mathbf{grad}_{g}\psi(x)\|_{g}^2}=\chi \boldsymbol{\nu} \qquad \chi=\dfrac{1}{\|\mathbf{grad}_{g}\psi(x)\|_{g}}.
\end{equation}

In order to pass the derivative into the integral in eq.\eqref{eq:derivativefunctionLevelset}
we use the transport property of integral under the action of the one-parameter group of diffeomorphisms:
\begin{equation}
\label{eq:Derivation_in_int}
\begin{split}
\dfrac{\D F}{\D v}(v)&=\lim_{s\rightarrow+\infty}\dfrac{\displaystyle{\int_{\mathrm{Fl}_s(\LevelSetfunc{v}{\psi})}\,f
\,\D\sigma_{\LevelSetfunc{v}{\psi},g}-\int_{\LevelSetfunc{v}{\psi}}\,f
\,\D\sigma_{\LevelSetfunc{v}{f},g}}}{s}=\lim_{s\rightarrow+\infty}\int_{\LevelSetfunc{v}{\psi}}\,\dfrac{\mathrm{Fl}^{*}_s(f
\,\D\sigma_{\LevelSetfunc{v}{\psi},g})-(f
\,\D\sigma_{\LevelSetfunc{v}{\psi},g})}{s}=\\
&=\int_{\LevelSetfunc{v}{\psi}}\,\mathcal{L}_{\boldsymbol{\xi}}(f\,\D\sigma_{\LevelSetfunc{v}{\psi},g})=\int_{\LevelSetfunc{v}{\psi}}\,\chi\mathcal{L}_{\boldsymbol{\nu}}(f\,\D\sigma_{\LevelSetfunc{v}{\psi},g})=\int_{\LevelSetfunc{v}{\psi}}\,\chi\left[\mathcal{L}_{\boldsymbol{\nu}}(f)\D\sigma_{\LevelSetfunc{v}{\psi},g}+f\mathcal{L}_{\boldsymbol{\nu}}(\D\sigma_{\LevelSetfunc{v}{\psi},g}))\right]
\end{split}
\end{equation}
where we have used the definition of the Lie derivative $\mathcal{L}_{\boldsymbol{\xi}}$ of forms with respect to the vector field $\boldsymbol{\xi}$, and we used its linearity with respect to reparametrization of
the one-parameter group of diffeomorphisms.
As 
\begin{equation}
\label{eq:derivationAreaForm}
\mathcal{L}_{\boldsymbol{\nu}}(\D\sigma_{\LevelSetfunc{v}{\psi},g})=\mathrm{Tr}^{g}(\mathrm{II})\D\sigma_{\LevelSetfunc{v}{\psi},g}=\tau_{1,g}\D\sigma_{\LevelSetfunc{v}{\psi},g}\,\, ,
\end{equation}
where $\mathrm{II}_{g}$ is the second fundamental form of the hypersurface $\LevelSetfunc{v}{\psi}$
and $\tau_{1,g}$ is the sum of principal curvatures (see \autoref{ch:DiffGeo}),
the last expression in eq.\eqref{eq:Derivation_in_int} can be rewritten as
\begin{equation}
\label{eq:Derivation_in_intII}
\dfrac{\D F}{\D v}(v)=\int_{\LevelSetfunc{v}{\psi}}\,\chi\left[\mathcal{L}_{\boldsymbol{\nu}}(f)+f\tau_{1,g}\right]\D\sigma_{\Sigma_{v}^N,g}\,.
\end{equation} 

In order to complete the proof it is sufficient to show that
$\mathrm{div}_{g}(f\boldsymbol{\nu})$ is equal to the expression in square brackets in eq.\eqref{eq:Derivation_in_intII}. Let us choos an adapted orthonormal frame
$(\boldsymbol{\nu},\boldsymbol{e}_1,...,\boldsymbol{e}_{N-1})$ to the regular set $\LevelSetfunc{v}{\psi}$, we have
\begin{equation}
\begin{split}
\mathrm{div}_{g}(f\boldsymbol{\nu})&=\sum_{i=1}^{N-1}g(\nabla_{\boldsymbol{e}_i}(f\boldsymbol{\nu}),\boldsymbol{e}_i)+g(\nabla_{\boldsymbol{\nu}}(f\boldsymbol{\nu}),\boldsymbol{\nu})=\\
&=\sum_{i=1}^{N-1}\,fg(\nabla_{\boldsymbol{e}_i}\boldsymbol{\nu},\boldsymbol{e}_i)+\sum_{i=1}^{N-1}\,(\nabla_{\boldsymbol{e}_i}f)\,g(\boldsymbol{\nu},\boldsymbol{e}_i)+fg(\nabla_{\boldsymbol{\nu}}\boldsymbol{\nu},\boldsymbol{\nu})+(\nabla_{\boldsymbol{\nu}}f)g(\boldsymbol{\nu},\boldsymbol{\nu})\,.
\end{split}
\end{equation}
Using the definition of the second fundamental form $\mathrm{II}_{g}(\mathbf{X},\mathbf{Y})=g(\nabla_{\mathbf{X}}\boldsymbol{\nu},\mathbf{Y})$ and the orthormality of the adapted frame we obtain:
\begin{equation}
\mathrm{div}_{g}(f\boldsymbol{\nu})=f\sum_{i=1}^{N-1}g(\nabla_{\boldsymbol{e}_i}\boldsymbol{\nu},\boldsymbol{e}_i)+(\nabla_{\boldsymbol{\nu}}f)=f\mathrm{Tr}^{g}(\mathrm{II}_{g})+\mathcal{L}_{\boldsymbol{\nu}}f=\mathcal{L}_{\boldsymbol{\nu}}f+f\tau_{1,g}
\end{equation}
as the actions of the covariant derivative and of the Lie derivative coincide on functions.

\begin{remark}
This results is implicitly contained in the geometrical microcanonical formalism developed
by Rugh in \cite{rugh1997dynamical,rugh1998geometric,rugh2001microthermodynamic} and Franzosi \cite{franzosi2011microcanonical}. Nevertheless we present this proof as we are interested to
stress the connection among the thermodynamics of a (configurational) microcanonical system and the \textbf{geometrical properties} related with the \textbf{Riemannian structure} of  configuration space.
\end{remark}

As a corollary of the theorem above we obtain the following results for Euclidean spaces:

\begin{corollary}[Federer,Laurence (\cite{federer2014geometric}\cite{Laurence1989})]

Let $O \subset \mathbb{R}^p$ be a bounded open set. Let $\psi \in {\cal C}^{n+1} (\overline O)$
be constant on each connected component of the boundary $\partial O$ and $f \in {\cal C}^n (O)$.
Define $O_{t,t'}=\{x \in O\mid t<\psi (x) <t' \}$ and

\begin{equation}
F(v)= \int_{ \{ \psi=v \} } f~\D\sigma^{p-1}
\end{equation} 
where $d\sigma^{p-1}$ represents the Lebesgue measure of dimension $p-1$.
If $~C>0$ exists such that for any $x \in O_{t,t'}, \Vert \boldsymbol{\mathrm{grad}}_{\mathbb{R}^p}
\psi (x) \Vert_{\mathbb{R}^p} \geq C$, then for any $k$ such that  $0 \leq k \leq n$,
for any  $v \in ]t,t'[$, one has
\begin{equation}
\label{eq: FedLau_Formula}
\frac{\D^k F}{\D v^k}(v) =\int_{\{ \psi=v \}} A_{\psi,\mathbb{R}^p}^k f~\D\sigma^{p-1} \ .
\end{equation}
with 
\begin{equation}
\label{eq: def_FedLauOperator}
A_{\psi,\mathbb{R}^p} f=\mathrm{div}_{\mathbb{R}^p} \left ( \dfrac{\boldsymbol{\mathrm{grad}}_{\mathbb{R}^p}
\psi}{\|\boldsymbol{\mathrm{grad}}_{\mathbb{R}^p}
\psi \|_{\mathbb{R}^p}} f \right )
\dfrac{1}{\| \boldsymbol{\mathrm{grad}}_{\mathbb{R}^p}
\psi \|_{\mathbb{R}^p}}
\end{equation}
\label{Cor:Federer}
\end{corollary}

\begin{remark}

The operator $A_{\psi,g}$ acting on the set of $\mathit{C}^{\infty}$ functions defined over the manifold $\mathcal{B}_{N}$ \textbf{is not} a derivation.
Although $A_{\psi,g}$ is $\mathbb{R}$-linear (as it is the sum of $\mathbb{R}$-linear operators), \textbf{it does not verify the Leibniz rule}, i.e.:
\begin{equation}
\begin{split}
A_{\psi,g}(fh)&=f_1 f_2 \chi\tau_{1,g}+f_2\mathcal{L}_{\boldsymbol{\xi}_N}(f_1)+f_1\mathcal{L}_{\boldsymbol{\xi}_N}(f_2)\chi\tau_{1,g}= f_1(f_2\chi\tau_{1,g}+\mathcal{L}_{\overline{\boldsymbol{\xi}}_N}f_2)+\\
&+f_2(f_1\chi\tau_{1,g}+\mathcal{L}_{\overline{\boldsymbol{\xi}}_N}f_1)-(f_1f_2)\chi\tau_{1,g}=\\
&=f_1 A_{\psi,g}(f_2)+f_2 A_{\psi,g}(f_1)-(f_1 f_2)\chi\tau_{1,g}\neq\\
&\neq f_1 A_{\psi,g}(f_2)+f_2 A_{\psi,g}(f_1)\,.
\end{split}
\end{equation}
for two arbitrary $C^{\infty}$ functions $f_1, f_2$ over $\mathcal{B}_N$
\end{remark}

\autoref{Th:GeneralizedFederer} allows also to calculate higher order derivatives of  the microcanonical partition function $\Omega_{n}(\bv)$ at any order.

\begin{corollary}[Higher order derivatives of the microcanonical partition function]

Let $O$ be an open bounded set of a $N$-dimensional Riemannian manifold $\Lambda_q,g$ and let  $\nabla$ be a Levi-Civita connection. Let $\overline{V}_N\in \mathcal{C}^{p+1}(\overline{O})$ be a generalized
potential constant on each connected component of the boundary $\partial\overline{O}$ and $f\in
\mathcal{C}^{p}(O)$. Define $M_{]\bv_0,\bv_1[}^N=\left\{x\in O |\bv_0<\overline{V}_N(x)<\bv_1\right\}$ 
and
\begin{equation}
\Omega_N(\bv)=\intSigmaV{\bv}{N} \,\D\mu_{\bv}^{N-1}
\end{equation}
where $\D\mu_{\bv}^{N-1}$ is the microcanonical  $(N-1)$-area-form of
eq.\eqref{eq:def_microcanonical_areaform} induced
over $\intSigmaV{\bv}{N}$.
If there exists $C>0$ such that for any $M_{]\bv_0,\bv_1[}^N$,
$\|\mathbf{grad}_{g}\overline{V}_N(x)\|_{g}\geq C$, and if the level sets
$\Sigma_{\bv}^N$ of $\overline{V}_N$ are without boundary, then for any $k$ such that $0\leq k\leq p $,
for any $\bv\in]\bv_0,\bv_1[$, one has
\begin{equation}
\label{eq:def_recursiveDer_Omega_microcan}
\dfrac{\D^k \Omega_N}{\D \overline{v}^k}(\bv)=\intSigmaV{\bv}{f}\,
A^{k}_{\mu}(1)\,\D\mu_{\bv}^{N-1}
\end{equation}
with
\begin{equation}
\label{eq:FedererOperator_microcan}
A_{\mu}(f)=f\mathrm{div}_{g}\left(\overline{\boldsymbol{\xi}}_N\right)+\overline{\chi}_N\mathcal{L}_{\boldsymbol{\nu}_N}(f)=f\overline{\zeta}_N+\mathcal{L}_{\overline{\boldsymbol{\xi}}_N}(f)
\end{equation}
\end{corollary}

\begin{remark}[Derivatives of $\Omega_{N}(\bv)$  and properties of diffeomorphisms of level sets] 
Equations \eqref{eq:def_recursiveDer_Omega_microcan} and \eqref{eq:FedererOperator_microcan} relate the  thermodynamic behaviour of the system considered (higher derivatives of microcanonical partition function) with the diffeomorphic properties of equipotential level sets through the
scalar quantities related to the vector field $\overline{\boldsymbol{\xi}}_N$: its divergence $\overline{\zeta}_N$ and its module $\overline{\chi}_N$.
This would lead to the conclusion that some suitable analytical constraint on the behaviour of $\overline{\zeta}_N$ and $\overline{\chi}_N$ can determine the absence of phase transitions in certain given family of level sets.
\end{remark}

Resorting to the above given formulas, we can readily express the derivatives of the microcanonical entropy as integrals of quantities over hypersurfaces only related to vector field $\overline{\boldsymbol{\xi}}_N$. 
Equations\eqref{eq:def_recursiveDer_Omega_microcan} and eqs.\eqref{eq:FedererOperator_microcan} allow to derive the links between microcanonical thermodynamics on one side and the geometrical properties of  the vector field $\overline{\boldsymbol{\xi}}_N$ on the other side.
The core of the proof of Necessity Theorem\cite{pettini2007geometry} consists in constructing uniform bounds in $N$ for the derivatives of configurational microcanonical entropy $\overline{S}_{N}(\bv)$ up to the fourth order: so we begin by calculating the derivatives of configurational microcanonical
partition function $\Omega_N(\bv)$ up to the fourth order with respect to $\bv$

\begin{equation}
\label{eq:Omega_Derivatives_Xi}
\begin{split}
&\dfrac{\D \Omega_N }{\D \bv}(\bv)=\intSigmaV{\bv}{\overline{V}_N}
\overline{\zeta}_N \,\D\mu^{N-1}_{\bv}\\
&\dfrac{\D^2 \Omega_N }{\D \bv^2}(\bv)=\intSigmaV{\bv}{\overline{V}_N}\left[
\overline{\zeta}_N^2+\mathcal{L}_{\overline{\boldsymbol{\xi}}_N}
(\overline{\zeta}_N)\right]\D\mu^{N-1}_{N\bv}\\
&\dfrac{\D^3 \Omega_N }{\D \bv^3}(\bv)=\intSigmaV{\bv}{\overline{V}_N}\left[
\overline{\zeta}_N^3+3\overline{\zeta}_N\mathcal{L}_{\overline{\boldsymbol{\xi}}_N}\left(\overline{\zeta}_N\right)+
\mathcal{L}^{(ii)}_{\overline{\boldsymbol{\xi}}_N}
(\overline{\zeta}_N)\right]\D\mu^{N-1}_{\bv}\\
&\dfrac{\D^4 \Omega_N
}{\D\overline{v}^4}(\overline{v})=\intSigmaV{\bv}{\overline{V}_N}\left[\overline{\zeta}_N^4+6\overline{\zeta}_N^2\mathcal{L}_{\overline{\boldsymbol{\xi}}_N}(\overline{\zeta}_N)+4\overline{\zeta}_N
\mathcal{L}^{(ii)}_{\overline{\xi}_N}(\overline{\zeta}_N)+3\left(\mathcal{L}_{\overline{\boldsymbol{\xi}}_N}(\overline{\zeta}_N)\right)^2+\mathcal{L}^{(iii)}_{\overline{\boldsymbol{\xi}}_N}(\overline{\zeta}_N)\right]\D\mu^{N-1}_{\bv}
\end{split}
\end{equation}

We recall that the configurational microcanonical entropy density is given by
\begin{equation}
\label{eq: MicroCan_Entropy_xi}
\overline{S}_{N}(\overline{v})=\dfrac{1}{N}\ln\Omega_{N}(\overline{v})=\dfrac{1}{N}\ln\intSigmaV{\bv}{\overline{V}_N}\,\mathrm{d}\mu^{N-1}_{\overline{v}}
\end{equation}
so its derivatives are given by:
\begin{equation}
\label{eq: microcanEntropy_Derivative}
\begin{split}
&\dfrac{\D \overline{S}_N}{\D
\overline{v}}(\overline{v})=\dfrac{1}{N}\dfrac{\Omega^{'}_N(\overline{v})}{\Omega_N(\overline{v})}\\
&\dfrac{\D^2 \overline{S}_N }{\D \overline{v}^2}(\overline{v})=\dfrac{1}{N}\,\left[
\dfrac{\Omega^{''}_N(\overline{v})}{\Omega_N(\overline{v})}-\left(\dfrac{\Omega^{'}_N(\overline{v})}{\Omega_N(\overline{v})}\right)^2 \right]\\
&\dfrac{\D^3 \overline{S}_N }{\D
\overline{v}^3}(\overline{v})=\dfrac{1}{N}\left[\dfrac{\Omega^{'''}_N(\overline{v})}{\Omega_N(\overline{v})}-3\dfrac{\Omega_N^{''}(\overline{v})}{\Omega_N(\overline{v})}
\dfrac{\Omega_N^{'}(\overline{v})}{\Omega_N(\overline{v})}+2\left(\dfrac{\Omega_N^{'}(\overline{v})}{\Omega_N(\overline{v})}\right)^3\right]\\
&\dfrac{\D^4 \overline{S}_N }{\D
\overline{v}^4}(\overline{v})=\dfrac{1}{N}\left[\dfrac{\Omega^{(iv)}_{N}(\overline{v})}{\Omega_N(\overline{v})}-4\dfrac{\Omega^{'''}_N(\overline{v})\Omega^{'}_N(\overline{v})}{\Omega_N^2(\overline{v})}+
12\dfrac{\Omega^{'2}_N(\overline{v})\Omega^{''}_N(\overline{v})}{\Omega_N^3(\overline{v})}-3\left(\dfrac{\Omega^{''}_N(\overline{v})}{\Omega_N(\overline{v})}\right)^2-
6\left(\dfrac{\Omega^{'}_N(\overline{v})}{\Omega_N(\overline{v})}\right)^4\right]\,.
\end{split}
\end{equation} 

To express also the derivatives of the microcanonical entropy density in terms of the scalar functions $\overline{\chi}_N$ and $\overline{\zeta}_N$, and of their Lie derivatives with respect to 
the vector field $\overline{\boldsymbol{\xi}}_N$, it is convenient to introduce the following notation for the average of a generic measurable function $f:M^{N}\to\mathbb{R}$ over the
hypersurface $\LevelSetfunc{\bv}{\overline{V}_N}$ endowed with microcanonical measure $\D\mu^{N-1}_{\bv}$.

\begin{equation}
\label{def:averageonSigmav}
\left\langle f \right\rangle_{\overline{v},\mu}=\dfrac{\displaystyle{\intSigmaV{\bv}{\overline{V}_N}\,f\D\mu_{\overline{v}}^{N-1}}}{\displaystyle{\intSigmaV{\bv}{\overline{V}_N}\,\D\mu_{\overline{v}}^{N-1}}}=
\dfrac{\displaystyle{\intSigmaV{\bv}{\overline{V}_N}\,f\D\mu_{\overline{v}}^{N-1}}}{\Omega_N(\overline{v})}\,\,.
\end{equation}
Consequently, we introduce the quantities
\begin{equation}
\label{def:defintition_statQuant}
\begin{split}
&\mathrm{Var}_{\overline{v},\mu}(f)=\mathrm{Cuml}^{(2)}_{\overline{v},\mu}(f)=\left\langle
f^2 \right\rangle_{\overline{v},\mu}-\left\langle f \right\rangle_{\overline{v},\mu}^2\\
&\mathrm{Corr}_{\overline{v},\mu}(f;g)=\left\langle f g
\right\rangle_{\overline{v},\mu}-\left\langle
f\right\rangle_{\overline{v},\mu}\left\langle g \right\rangle_{\overline{v},\mu}\\
&\mathrm{Cuml}^{(3)}_{\overline{v},\mu}(f)=\left\langle f^3
\right\rangle_{\overline{v},\mu}-3\left\langle f
\right\rangle_{\overline{v},\mu}\left\langle f^2
\right\rangle_{\overline{v},\mu}+2 \left\langle f \right\rangle_{\overline{v},\mu}^3\\
&\mathrm{Cuml}^{(4)}_{\overline{v},\mu}(f)=\left\langle f^4 \right\rangle_{\overline{v},\mu}-4\left\langle f^3 \right\rangle_{\overline{v},\mu}\left\langle f
\right\rangle_{\overline{v},\mu}+12\left\langle f^2\right\rangle_{\overline{v},\mu}\left\langle f\right\rangle_{\overline{v},\mu}^2-3\left\langle f^2\right\rangle_{\overline{v},\mu}^2-6\left\langle f\right\rangle_{\overline{v},\mu}^4\\
\end{split}
\end{equation}
which represent the variance, the correlation function, and the 3rd and 4th order 
cumulants on the hypersurface $\LevelSetfunc{\bv}{\overline{V}_N}$ with measure
$\D\mu^{N-1}_{\bv}$, respectively. 

With this notation and substituting eqs.\eqref{eq:Omega_Derivatives_Xi} in eqs.\eqref{eq: microcanEntropy_Derivative} it is possible to show that the derivatives of the microcanonical
entropy at a non critical value $\bv$, and at fixed $N$, can be tightly related to the vector field $\overline{\boldsymbol{\xi}}_N$ which generates the diffeomorphisms among the equipotential level sets:

\begin{equation}
\begin{split}
&\dfrac{\D \overline{S}_N}{\D
\overline{v}}(\overline{v})=\dfrac{1}{N}\left\langle\overline{\zeta}_N\right\rangle_{\overline{v},\mu}\\
&\dfrac{\D^2 \overline{S}_N }{\D
\overline{v}^2}(\overline{v})=\dfrac{1}{N}\left[\mathrm{Var}_{\overline{v},\mu}(\overline{\zeta}_N)+\left\langle\mathcal{L}_{\overline{\boldsymbol{\xi}}_N}
(\overline{\zeta}_N)\right\rangle_{N\bv,\mu}\right]\\
&\dfrac{\D^3 \overline{S}_N }{\D\overline{v}^3}(\bv)=\dfrac{1}{N}\left[\mathrm{Cuml}^{(3)}_{\overline{v},\mu}(\overline{\zeta}_N)+3\mathrm{Corr}_{\overline{v},\mu}\left(\overline{\zeta}_N;\mathcal{L}_{\overline{\boldsymbol{\xi}}_{N}}(\overline{\zeta}_N)\right)+\left\langle\mathcal{L}_{\overline{\boldsymbol{\xi}}_N}^{(ii)}\left(\overline{\zeta}_N\right)\right\rangle_{\overline{v},\mu}\right]\\
&\dfrac{\D^4 \overline{S}_N }{\D\overline{v}^4}(\bv)=\dfrac{1}{N}\Biggr[\mathrm{Cuml}^{(4)}_{\overline{v},\mu}(\overline{\zeta}_N)+6\mathrm{Corr}_{\overline{v},\mu}\left(\overline{\zeta}_N^2;\mathcal{L}_{\overline{\boldsymbol{\boldsymbol{\xi}}}_N}(\overline{\zeta}_N)\right)+
3\mathrm{Var}_{\overline{v},\mu}\left(\mathcal{L}_{\overline{\boldsymbol{\boldsymbol{\xi}}}_N}(\overline{\zeta}_N)\right)+\\
&+4\mathrm{Corr}_{\overline{v},\mu}\left(\overline{\zeta}_N;\mathcal{L}_{\overline{\boldsymbol{\boldsymbol{\xi}}}_N}^{(ii)}(\overline{\zeta}_N)\right)
-12\left\langle\overline{\zeta}_N\right\rangle_{\overline{v},\mu}\mathrm{Corr}_{N\overline{v},\mu}\left(\overline{\zeta}_N;\mathcal{L}_{\overline{\boldsymbol{\boldsymbol{\xi}}}_N}(\overline{\zeta}_N)\right)+\left\langle\mathcal{L}_{\overline{\boldsymbol{\boldsymbol{\xi}}}_N}^{(iii)}\left(\overline{\zeta}_N\right)\right\rangle_{\overline{v},\mu}\Biggr]=\\
&=\dfrac{1}{N}\Biggr[\mathrm{Cuml}^{(4)}_{\overline{v},\mu}(\overline{\zeta}_N)+4\mathrm{Corr}_{\overline{v},\mu}\left(\overline{\zeta}_N;\mathcal{L}_{\overline{\boldsymbol{\boldsymbol{\xi}}}_N}^{(ii)}(\overline{\zeta}_N)\right)+3\mathrm{Var}_{\overline{v},\mu}\left(\mathcal{L}_{\overline{\boldsymbol{\boldsymbol{\xi}}}_N}(\overline{\zeta}_N)\right)+\\
&+6\left\langle\overline{\zeta}_N\right\rangle_{\overline{v},\mu}\left(\mathrm{Corr}_{\overline{v},\mu}\left(\Delta\overline{\zeta}_N;\mathcal{L}_{\overline{\boldsymbol{\boldsymbol{\xi}}}_N}(\overline{\zeta}_N)\right)\right)+\left\langle\mathcal{L}_{\overline{\boldsymbol{\boldsymbol{\xi}}}_N}^{(iii)}\left(\overline{\zeta}_N\right)\right\rangle_{\overline{v},\mu}
\Biggr]
\end{split}
\label{eq:microcanEntropy_Derivative_Xi}
\end{equation} 
where for sake of simplicity we have introduced the quantity
\begin{equation}
\Delta \overline{\zeta}_N=\dfrac{\overline{\zeta}_N^2}{\left\langle \overline{\zeta}_N \right\rangle_{\overline{v},\mu}}-2\overline{\zeta}_N \,\,\,.
\end{equation}

As mentioned above, the first important consequence that can be argued by eqs.\eqref{eq:microcanEntropy_Derivative_Xi} is that, in principle, it is possible to directly control  the behaviour of microcanonical entropy and its derivatives at any finite $N$ and in the
thermodynamic limit.
This is obtained by imposing some conditions on the behaviour of the components of the vector $\overline{\boldsymbol{\xi}}_N$.\\
This result opens the possibility to refine the Pettini-Franzosi Theorem. In fact, the requirement of diffeomorphicity among the equipotential level sets at \textit{any finite} $N$ - in a given interval of $\bv$ values - is not sufficient to avoid the occurrence of a phase transition in the \textit{thermodynamic limit} in the same interval of specific potential energy values.
Thanks to eqs.\eqref{eq:microcanEntropy_Derivative_Xi} it is possible to control \textit{asymptotically} in $N$ ``how'' and/or ``how much'' the level sets have to be diffeomorphic among themselves in order to prevent  the occurrence of phase transitions. This is shown in what follows.

\section{Geometrization of thermodynamics through regular equipotential level sets}
Eqs.\eqref{eq:microcanEntropy_Derivative_Xi} open to the possibility of directly associating the microcanonical entropy and its derivatives to geometrical 
features of regular equipotential level sets. In fact, using the notation introduced
in this Section
\begin{equation}
\label{eq:GeomInt_DivXi}
\zeta_N
(p)=\mathrm{div}_{\mathbb{R}^N}(\overline{\boldsymbol{\xi}}_N)(p)=\mathrm{div}_{\mathbb{R}^N}\left(\overline{\chi}_N\boldsymbol{\nu}_N\right)(p)=\overline{\chi}_N(p)h_{1,g_{\mathbb{R}^N}}(p)+(\mathcal{L}_{\boldsymbol{\nu}_N}\overline{\chi}_N)(p)
\end{equation}
where $p\in\LevelSetfunc{\bv}{\overline{V}_N}\subset\mathbb{R}^N$ and $h_{1,g_{\mathbb{R}^N}}$ is the
sum of principal curvatures of $\LevelSetfunc{\bv}{\overline{V}_N}$ immersed in $\mathbb{R}^N$ endowed with the metric induced by the ambient space (see \autoref{ch:DiffGeo}).\\
Substituting the above expression in the first of Eqs.\eqref{eq:Omega_Derivatives_Xi}
\begin{equation}
\label{eq:Omega_DerivativeI_Xibis}
\begin{split}
\dfrac{\D \Omega_N}{\D \overline{v}}(\overline{v})=&N\,\int_{\Sigma^N_{N\overline{v}}}
\left[-\chi_N h_{1,g_{\mathbb{R}^N}}+(\mathcal{L}_{\nu_N}\chi_N)
\right]\D\mu^{N-1}_{N \overline{v}}=\\
=&N\,\int_{\Sigma^N_{N\overline{v}}}
\left[-\chi_N (h_{1,g_{\mathbb{R}^N}}+(\mathcal{L}_{\nu_N}(-\log\chi_N))
\right]\exp\left[-\log\chi_N\right]\D\sigma^{N-1}_{N
\overline{v}}
\end{split}
\end{equation}
where the last expression has been derived to compare
this result with well known results in the theory of manifolds with density.

\subsection{Regular equipotential surfaces as manifolds with density}
\label{subsec:ManWDens}
The derivatives of configurational microcanonical partition function with
respect to $\bv$ are reported in eq.\eqref{eq:Omega_Derivatives_Xi} \textit{only}
as functions of integral quantities of $\overline{\zeta}_N$ and its Lie derivatives
with respect to the vector field $\overline{\boldsymbol{\xi}}_N$. This establishes
a strong relation between configurational microcanonical thermodynamics and 
diffeomorphic properties of the level sets; this was motivated by the search for a
proper framework to define the concept of "asymptotic change of topology". Another
possible approach to this problem would consist to interpret the integral quantities
that enter eqs.\eqref{eq:Omega_Derivatives_Xi} in terms of "geometrical observables"
(especially curvatures) of the equipotential level sets and of the (configuration)
ambient space. Eq.\eqref{eq:GeomInt_DivXi} suggests a link between $\overline{\zeta}_N$
and the mean curvature of equipotential level sets. 
Nevertheless, at this level it is not clear how to give a pure geometrical interpretation
of the integrals in eqs. \eqref{eq:Omega_Derivatives_Xi}. To attain this result,
we consider quite recent results on the differential geometry of \textit{manifolds with density}\footnote{This objects are widely studied in the context of isoperimetric problems and in optimal transportation theory.} \cite{morgan2005manifolds}\cite{corwin2006differential}.
In this framework, usual geometric quantities (as curvatures) of a Riemannian manifold are redefined in order to encode in the geometry also the information carried by an arbitrary measure
over a manifold.\\
Let $M$ be a Riemannian manifold endowed with a metric $g$, and consider an immersed codimension
one submanifold $\Sigma\subset M$; the Riemannian volume forms $\D \mathrm{Vol}_{g}$ and
$\D\sigma_{\Sigma,g}$ are induced on the manifold $M$ and $\Sigma$ respectively. \\
To construct a manifold with density, a density function $\Psi:M\longrightarrow\mathbb{R}^{+}$ is defined on $M$ so that the volume form and the area form on $\Sigma$ are rescaled in order to give respectively 
\begin{equation}
\begin{split}
&\D\mathrm{Vol}_{\Psi}=\Psi \D\mathrm{Vol}_{g}=\exp[\psi]\D\mathrm{Vol}_{g}\\
&\D\sigma_{\Sigma,\Psi}=\Psi\D\sigma_{\Sigma,g}=\Psi \D\mathrm{Vol}_{g}\Bigr|_{S}=\exp[\psi]\D\mathrm{Vol}_{g}\Bigr|_{\Sigma}\, .\\
\end{split}
\end{equation}
where $\psi=\ln\Psi$.\\
The mean curvature $h_{1,g}=\tau_{1,g}/(N-1)$ of the hypersurface $\Sigma$ is redefined with the introduction of the density such that the sum of principal curvatures $\tau_{1,g,\Psi}$ is directly proportional to the variation of area element $\D\sigma_{\Sigma,\Psi}$ at the first order in normal direction to the level set hypersurfaces, i.e.
\begin{equation}
\mathcal{L}_{\boldsymbol{\nu}}\D\sigma_{\Sigma,\Psi}=-\tau_{1,g,\Psi}\D\sigma_{\Sigma,\Psi}=-(N-1)h_{(1,g,\Psi)}\D\sigma_{\Sigma,\Psi}
\end{equation}
which is immediately verified.\\
For a manifold with density $(M,g,\Psi)$ a natural extension of the sum of principal curvatures $\tau_{1,g}$ is thus given by:
\begin{equation}
\tau_{(1,g,\Psi)}=\tau_{1,g}+\mathcal{L}_{\boldsymbol{\nu}}\psi
\end{equation}
where $\boldsymbol{\nu}_N$ is the normal vector field to the hypersurface.
 Consistently the first and second variation formula for the area of the hypersurface $S$ are
\begin{equation}
\mathcal{A}_{S}=\int_{S}\,\D\sigma_{\Psi}
\end{equation}
under the action of a diffeomorphism parametrized by $t$ and generated by a vector field $W=w\nu$, such that $\D t(W)=1$, we have \cite{bayle2003proprietes} 
\begin{equation}
\label{eq:ManwDens_FirstAreaVar}
\dfrac{\D \mathcal{A}_{S}}{\D t}=\int_{S}\, -w \,h_{(1,g,\Psi)} \D\sigma_{\Psi}
=\int_{S}\,- w\left(h_{1,g}+\mathcal{L}_{\nu}\psi\right)\,\exp[\psi]\D\sigma_{g}
\end{equation}
and
\begin{equation}
\label{eq:ManwDens_SecondAreaVar}
\dfrac{\D^2 \mathcal{A}_{S}}{\D t^2}=\int_{S}\,\left[\|\mathbf{grad}_{g}w\|^{2}_{g}-w^2\left(h_{1,g}^2+\mathcal{L}_{\mathcal{\nu}_N}(\mathcal{L}_{\boldsymbol{\nu}}\psi)-\tau_{2,g}-\mathrm{Ric}_{g}(\boldsymbol{\mathcal{\nu}_N},\boldsymbol{\nu}_N)\right)\right]\,\exp[\psi]\D\sigma_{g}
\end{equation}

where $\tau_{2,g}=\|\mathrm{II}_g\|^2$ is the squared norm of the shape operator, i.e. it is the the sum of the squares of principal curvatures, and $\mathrm{Ric}_{g}$ is the Ricci curvature of the ambient space with metric $g$.\\
We easily see that with the identification $w=\overline{\chi}_N$, $\psi=-\log(\overline{\chi}_N)$ and,   consequently, $\Psi=\overline{\chi}_N^{-1}$ we exactly  obtain the expression in eqs.\eqref{eq:Omega_DerivativeI_Xibis}. 

\subsection{Rescaled metric in configuration space}

In \autoref{subsec:ManWDens} we have discussed how it is possible to interpret equipotential level
sets equipped with the microcanonical measure as manifolds with density, "geometrizing" some features strictly related with measure properties.\\
Nevertheless, in an ideal program of "geometrization" of classical microcanonical thermodynamics", \textit{all the terms} in the integrands of eqs.\eqref{eq:Omega_Derivatives_Xi} should be retrieved only from the geometrical and topological properties of the equipotential level sets and of the ambient space. In particular, according to what has been reported in the previous Subsection, the function $\overline{\chi}_N$ (well defined in absence of critical points of potential energy) and its derivatives in the normal direction $\boldsymbol{\nu}_N$ carry two distinct information that have to be "geometrized": one concerns the density measure $\Psi=e^{\overline{\chi}_N}$, while the other concerns the "velocity" of the vector field that "moves" the level set $w=\overline{\chi}_N$. A possible way consists in the introduction of a rescaled metric $\tilde{g}$ where the microcanonical measure $\D \mu_{\LevelSetfunc{\bv}{\overline{V}_N}}$ and the vector field $\overline{\boldsymbol{\xi}}_N$ are "natural" in the sense that they are naturally included in the differential geometrical structure of the space.\\
In the following section we introduce the function $\phi_{N}:\mathcal{B}_N\subseteq\mathcal{X}_N\longrightarrow\mathbb{R}_{0}^{+}$ such that:
\begin{equation}
\label{eq:def_phifunction}
\phi_{N}=\log\overline{\chi}_N
\end{equation}
to simplify the notation.\\
Although not strictly necessary, for further computations it is convenient to introduce a coordinate system $\{u^0,u^1,...,u^{N-1}\}$ over $M_{[\bv_0,\bv_1]}\subset\mathcal{B}_N$ such
that one coordinate parametrizes the specific potential energy
\begin{equation}
\D u^0=\D \overline{V}_N \quad \Longrightarrow \quad \D u^0(\overline{\boldsymbol{\xi}}_N)=1
\end{equation}
$\{\boldsymbol{\partial}_0,\boldsymbol{\partial}_1,...,\boldsymbol{\partial}_{(N-1)}\}$ is the coordinate frame and $\{\D u_0,\D u_1,\ldots\D u_{(N-1)}\}$ its dual. The greek indices\footnote{Einstein's convention is assumed for repeated indices.} run in the interval $[0;N-1]$ while the latin indices refer to the coordinate system over the level set hypersurfaces and run in the interval $[1;N-1]$.\\
With this coordinate choice the $g$ metric of the ambient space reads
\begin{equation}
g=\sum_{\alpha=0}^{N-1}\sum_{\beta=0}^{N-1}g_{\alpha\beta}\D
u^{\alpha} \otimes \D u^{\beta}=e^{2\phi_N} \D u^0  \otimes \D u^0 +\sum_{i=1}^{N-1}\sum_{j=1}^{N-1} g_{ij}\D u^{i}\otimes \D u^{j}
\end{equation}
and the normal vector field $\overline{\boldsymbol{\nu}}_N$ defining the foliation of $M_{[\bv_0,\bv_1]}$ is 
\begin{equation}
\boldsymbol{\partial}_0 =\overline{\boldsymbol{\xi}}_N \quad \Longrightarrow \quad \boldsymbol{\nu}_N=e^{-\phi_N}\boldsymbol{\partial}_0
\end{equation}
With such a choice of coordinates, the Riemannian volume form of the ambient space and the area form induced over a fixed $\LevelSetfunc{\bv}{\overline{V}_N}\in M_{[\bv_0,\bv_1]}$ are respectively
\begin{equation}
\D \mathrm{Vol}_{g}=|\mathrm{det}(g_{\alpha,\beta})|\D u^0\wedge...\D u^{N-1}=e^{\phi_N}\left |\mathrm{det}(g_{ij})\right|\D u^0 \wedge \wedge \D u^1\wedge....\D u^{N-1}
\end{equation}
and 
\begin{equation}
\D \sigma_{\LevelSetfunc{\bv}{\overline{V}_N},g}=|\mathrm{det}(g_{ij})| \D u^1\wedge...\wedge\D u^{N-1}\,.
\end{equation}
Christoffel symbols $\Christoffel{\alpha}{\beta}{\gamma}$ of the Levi-Civita connection $\nabla$ associated with $g$ are supposed to be given.\\
As reported in \autoref{sec:geometry_regularlevelsets}, all the informations
concerning the statistical mechanics of the configurational microcananical ensemble
are given by functional depending on $\overline{\zeta}_N=\mathrm{div}_{g}\left(\overline{\boldsymbol{\xi}}_{N}\right)$ and its Lie derivatives respect to the same vector field $\overline{\boldsymbol{\xi}_N}$.
Eq.\eqref{eq:GeomInt_DivXi} clearly shows that the mean curvature of $\LevelSetfunc{\bv}{\overline{V}_N}$ and its derivatives with respect to $\boldsymbol{\nu}_N$ are the geometrical quantities of the level sets directly related with the derivatives of configurational microcanonical entropy.
For these reasons we provide an explicit expression of the mean curvature and its first order Lie derivatives respect to the normal vector field as a function of Christoffel of Levi-Civita connection of metric $g$.
With the choice of coordinates introduced at the beginning of this section, the sum of principle curvatures $\tau_{1,g}$ of a regular level sets $\LevelSetfunc{\bv}{\overline{V}_N}$
is given by:
\begin{equation}
\label{eq:sum_principalcurvature}
\begin{split}
\tau_{1,g}&\equiv\sum_{i,j=1}^{N-1}\lambda_{i,g}=\sum_{i,j=1}^{N-1}\mathrm{II}_{ij} g^{ij}=\sum_{i,j=1}^{N-1}
g(\nabla_{\partial_i}\nu_N,\partial_j)g^{ij}=\\
&=\sum_{i,l=1}^{N-1}g(\nabla_{\partial_i}\left(e^{-\varphi_N}\partial_{0}\right),\partial_j)g^{ij}=
e^{-\varphi_N}\sum_{i,l=1}^{N-1}\Christoffel{k}{i}{0}g_{kj}g^{ij}=e^{-\varphi_N} \Christoffel{i}{0}{i}\,.
\end{split}
\end{equation}
where $\mathrm{II}$ is the second fundamental form on the equipotential level sets
(see \autoref{ch:DiffGeo} for a brief review on differential geometry.).\\
Moreover, as a consequence of the Riccati's Equation applied to Weingarten operator
under the action of the vector field $\boldsymbol{\nu}_N$, we obtain:
\begin{equation}
\mathcal{L}_{\boldsymbol{\nu}_N}(\tau_{1,g})=-\tau_{2,g}-\mathrm{Ric}(\boldsymbol{\nu}_N,\boldsymbol{\nu}_N)
\end{equation}
where $\tau_{2,g}=\sum_{i=1}^{N-1}\lambda_{i,g}^2$ is the sum of the squares of principal curvatures and $\mathrm{Ric}$ is the Ricci tensor of the ambient space. Using the coordinate system introduced above we obtain using definitions:
\begin{equation}
\begin{split}
\mathrm{Ric}(\partial_0,\partial_0)&=\sum_{\alpha=0}^{N-1}g(R(\partial_{\alpha},\partial_0)\partial_{0},\partial_{\alpha})=\partial_{\alpha}\Christoffel{\alpha}{0}{0}-\partial_{0}\Christoffel{\alpha}{0}{\alpha}+\Christoffel{\alpha}{\alpha}{\beta}\Christoffel{\beta}{0}{0}-\Christoffel{\alpha}{0}{\beta}\Christoffel{\beta}{\alpha}{0}=\\
&=\partial_{i}\Christoffel{i}{0}{0}-\partial_{0}\Christoffel{i}{i}{0}+\Christoffel{i}{i}{0}\Christoffel{0}{0}{0}-\Christoffel{0}{0}{i}\Christoffel{i}{0}{0}-\Christoffel{i}{0}{j}\Christoffel{j}{i}{0}\\
\end{split}
\end{equation}
\begin{equation}
\begin{split}
\mathrm{Ric}(\nu_N,\nu_N)&=\mathrm{Ric}(e^{-\varphi_N}\partial_{0},e^{-\varphi_N}\partial_{0})=e^{-2\varphi_N}R_{00}=\\
&=e^{-2\varphi_N}\left(\partial_{i}\Christoffel{i}{0}{0}-\partial_{0}\Christoffel{i}{i}{0}+\Christoffel{i}{i}{0}\Christoffel{0}{0}{0}-\Christoffel{0}{0}{i}\Christoffel{i}{0}{0}-\Christoffel{i}{0}{j}\Christoffel{j}{i}{0}\right)\\
\end{split}
\end{equation}
while for the sum of the squares of principle curvatures is given by
\begin{equation}
\begin{split}
\tau_{2,g}&=\mathrm{Tr}^{g}\left(W^2\right)=\mathrm{II}_{ij}\mathrm{II}_{kl}g^{jk}g^{il}=g\left(\nabla_{\partial_i}(e^{-\varphi_N}\partial_{0}),\partial_{j}\right)g\left(e^{-\varphi_N}\partial_{0}),\partial_{l}\right)g^{jk}g^{il}=\\
&=e^{-2\varphi_N}g_{mj}g_{nl}\Christoffel{m}{i}{0}\Christoffel{n}{k}{0}g^{jk}g^{il}=e^{-2\varphi_N}\Christoffel{k}{i}{0}\Christoffel{i}{k}{0}
\end{split}
\end{equation}

\noindent As anticipated, we endow the manifold $M^N_{[\bv_0,\bv_1]}$ in configuration space with a new metric
$\tilde{g}$ satisfying the following properties:
\begin{enumerate}
\item  \label{itm:rescmetric_AreaHyp} 
the Riemannian area form $\D \sigma_{\tilde{g}}$ induced over the equipotential level sets $\Sigma_{\bv}^N \in M^N_{[\bv_0,\bv_1]}$ from ambient space \textit{would exactly correspond with the microcanonical density form}
\begin{equation}
\D \mu^{N-1}_{\bv}=\D\tilde{\sigma}_{\LevelSetfunc{\bv}{\overline{V}_N},\tilde{g}}=|\mathrm{det}(\tilde{g}_{ij})|\D u^0\wedge \D u^1\wedge...\wedge \D u^{N-1}\,;
\end{equation}
\item
 \label{itm:rescmetric_Xinorm} 
 the vector field $\overline{\boldsymbol{\xi}}_N\in\mathfrak{X}(M^N_{[\bv_0,\bv_1]})$ coincides with the 
\textit{normal vector to the equipotential hypersurfaces}
\begin{equation}
\label{eq:rescmetr_Normxicondition}
\tilde{g}(\overline{\boldsymbol{\xi}}_N,\overline{\boldsymbol{\xi}}_N)=1 \quad \Longrightarrow \quad \tilde{\boldsymbol{\nu}}_N=\overline{\boldsymbol{\xi}}_N \,.
\end{equation}
in this way the derivation with respect to the parameter $\bv$ (along the flow generated by $\overline{\boldsymbol{\xi}}_N$) coincides with the Lie derivative along the vector field $\tilde{\boldsymbol{\nu}}_N$.
\end{enumerate}
We notice that the first condition concerns the properties of the metric restricted to the \textit{tangent space of hypersurfaces} $\LevelSetfunc{\bv}{\overline{V}_N}$ while the second condition concerns a rescaling in \textit{normal direction}.\\
This suggests that a possible choice for $\tilde{g}$ can be done by performing two different conformal rescalings for the components of the metric $g$, i.e. the tangent and normal ones to the equipotential level sets, as follows:
\begin{equation}
\label{eq:rescaled_metric}
\begin{cases}
&\tilde{g}(\boldsymbol{\nu}_N,\boldsymbol{\nu}_N)=e^{-2\phi_N}
g(\boldsymbol{\nu}_N,\boldsymbol{\nu}_N)
\\
&\tilde{g}(\boldsymbol{X},\boldsymbol{Y})=e^{2\Xi\phi_N}
g(\boldsymbol{X},\boldsymbol{Y})\qquad\text{with}\qquad \Xi=\dfrac{1}{N-1} \\
&\tilde{g}(\boldsymbol{X},\boldsymbol{\nu}_N)=g(\boldsymbol{X},\boldsymbol{\nu}_N)=0
\end{cases}
\end{equation}
where $\boldsymbol{\nu}_N$ is the normal vector field to the equipotential level sets $ \LevelSetfunc{\bv}{\overline{V}_N}$ with respect to the metric $g$, and $\boldsymbol{X},\boldsymbol{Y}\in T_{p}\LevelSetfunc{\bv}{\overline{V}_N}$ are vector fields belonging to the tangent bundle of a leaf $ \LevelSetfunc{\bv}{\overline{V}_N}\in \Mfunc{[\bv_0,\bv_1]}{\overline{V}_N}$.\\

\begin{remark}[Restrictions to the definition of the rescaled metric $\tilde{g}$]
It has to be stressed that the suggested rescaling change of metric on $\Mfunc{[\bv_0,\bv_1]}{\overline{V}_N}$ is possible only in the case of absence of critical points of the specific potential energy $\overline{V}_N$, i.e. when the function $\overline{\chi}_N =\|\mathbf{grad}_{g}\overline{V}_N\|_{g}^{-1}>0$ is non singular. In this case the rescaled metric $\tilde{g}$ defined in\eqref{eq:rescaled_metric} is well defined and it is positive definite so that  $(\Mfunc{[\bv_0,\bv_1]}{\overline{V}_N},\tilde{g})$ is a Riemannian manifold. As the specific potential energy is 
in the closure of the Morse function set in $\Mfunc{[\bv_0,\bv_1]}{\overline{V}_N}$ for a large class of potentials. This implies that \textbf{the proposed rescaling of the metric $g$ for the geometrization of microcanonical thermodynamics is possible only under the hypothesis of diffeomorphicity of equipotential level sets $ \LevelSetfunc{\bv}{\overline{V}_N} \in \Mfunc{[\bv_0,\bv_1]}{\overline{V}_N}$ at any finite $N$}
\end{remark}

Using the local coordinate system  $\{u_{\alpha}\}_{\alpha=0,...,(N-1)}$ introduced at the beginning of this section, the rescaled metric $\tilde{g}$ reads
\begin{equation}
\label{eq:rescmetric_Def_coordinates}
\begin{split}
 \tilde{g}&=\sum_{\alpha=0}^{N-1}\sum_{\beta=0}^{N-1}\tilde{g}_{\alpha\beta}\D u^{\alpha} \D u^{\beta}=e^{-2\varphi_N} g_{00}\D u^0  \otimes \D u^0 +\sum_{i=1}^{N-1}\sum_{j=1}^{N-1}e^{2\Xi\phi_N}g_{ij}\D u^{i}\otimes \D u^{j}=\\
 &=\tilde{g}_{00}\,\,\D u^0  \otimes \D u^0 +\sum_{i=1}^{N-1}\sum_{j=1}^{N-1} \tilde{g}_{ij}\,\,\D u^{i}\otimes \D u^{j} \,.
 \end{split}
\end{equation}
With this rescaling of the metric it is quite simple to verify both condition (\autoref{itm:rescmetric_AreaHyp})
\begin{equation}
\begin{split}
\D \tilde{\sigma}_{\LevelSetfunc{\bv}{\overline{V}_N},\tilde{g}}&=|\mathrm{det}(\tilde{g}_{ij})|\D
u^1\wedge...\wedge \D u^{N-1}=\left(\prod_{i}^{N-1}\overline e^{2\Xi\phi_N}\right)^{1/2}
|\mathrm{det}(g_{ij})| \D u^1\wedge...\wedge\D u^{N-1}=\\
&=e^{\phi_N}\ |\mathrm{det}(g_{ij})| \D u^1\wedge...\wedge\D u^{N-1}=\overline{\chi}_N\D \sigma_{\LevelSetfunc{\bv}{\overline{V}_N},g}=\D \mu_{\LevelSetfunc{\bv}{\overline{V}_N}}
\end{split}
\end{equation}
and condition (\autoref{itm:rescmetric_Xinorm})
\begin{equation}
\tilde{g}(\overline{\boldsymbol{\xi}}_N,\overline{\boldsymbol{\xi}}_N)=\D u^0(\overline{\boldsymbol{\xi}}_N)\otimes \D u^0(\overline{\boldsymbol{\xi}}_N)=1 \quad \Longrightarrow \quad \tilde{\boldsymbol{\nu}}_N=\overline{\boldsymbol{\xi}}_N(=\boldsymbol{\partial}_0)
\end{equation}
in last equation we have used \eqref{eq:def_HirschVector}.

\begin{remark}[Preservation of ambient space volume]
The new metric introduced in eqs.\eqref{eq:rescaled_metric} preserves the Riemannian volume form $\D\mathrm{Vol}_{g}$,
in fact
\begin{equation}
\begin{split}
&\D \mathrm{Vol}_{\tilde{g}}=|\tilde{g}(\alpha,\beta)|\D u^{0}\wedge \D u^{1}\wedge...\wedge \D u^{N-1}=\left(\prod_{i}^{N-1}e^{2\Xi\varphi_N}\right)^{1/2} |\mathrm{det}(g_{ij})| \D u^0 \wedge...\wedge\D u^{N-1}=\\
&=e^{\varphi_N}\mathrm{det}(g_{ij})| \D u^{0}\wedge...\wedge\D u^{N-1}=\overline{\chi}_N |\mathrm{det}(g_{ij})| \D u^{0}\wedge...\wedge\D u^{N-1}=\D \mathrm{Vol}_{g}
\end{split}
\end{equation}
From the point of view of thermodynamic properties of the system, this means that the Gibbs' microcanonical volume  $\Omega_{\mathrm{Gibbs},N}$ and the Gibbs' microcanonical entropy
$\overline{S}_{Gibbs,N}=N^{-1}\ln\Omega_{\mathrm{Gibbs},N}$ are invariant for the transformation 
of the metric in eq.\eqref{eq:rescaled_metric}.
\end{remark}
The introduction of the rescaled metric $\tilde{g}$ allows to express the derivatives of configurational microcanonical entropy in terms of geometric properties of $\LevelSetfunc{\bv}{\overline{V}_N}$.\\
The microcanonical partition function $\Omega_N (\overline{v})$ becomes simply the Riemannian area of the hypersurfaces $\LevelSetfunc{\bv}{\overline{V}_N}$
\begin{equation}
\label{eq:rescMetr_Omega}
\Omega_N(\overline{v})=\int_{\LevelSetfunc{\bv}{\overline{V}_N}}\,\D
\tilde{\sigma}_{\LevelSetfunc{\bv}{\overline{V}_N},\tilde{g}}\,.
\end{equation}
Moreover, as required, the vector field that generates the one-parameter group  of diffeomorfisms  $\mathrm{Fl}_{\bv}$ among equipotential level sets coincides with the normal vector field $\boldsymbol{\tilde{\nu}}_N$.\\
According to eq.\eqref{eq:derivationAreaForm}, the Lie derivative along the vector field that generate the diffeomorphism $\tilde{\boldsymbol{\nu}}_N$ of the area form  reads in rescaled metric $\tilde{g}$
\begin{equation}
\label{eq:derivationAreaForm}
\mathcal{L}_{\overline{\boldsymbol{\nu}}_N}(\D\tilde{\sigma}_{\LevelSetfunc{\bv}{\overline{V}_N},\tilde{g}})=\mathrm{Tr}^{\tilde{g}}(\mathrm{II_{\tilde{g}}})\D\tilde{\sigma}_{\LevelSetfunc{\bv}{\overline{V}_N},g}=\tau_{1,\tilde{g}}\D\tilde{\sigma}_{\LevelSetfunc{\bv}{\overline{V}_N},\tilde{g}}\,\, ,
\end{equation}
our rescaling is consistent with our purposes of "geometrizating" the configurational microcanonical thermodynamics if $\tau_{1,\tilde{g}}=\mathrm{div}_{g}\overline{\boldsymbol{\xi}}_N$.\\
In the next Subsection a characterization of some geometrical properties of the Riemannian manifolds $(\Mfunc{\bv_0,\bv_1}{\overline{V}_N},\tilde{g})$ and of the equipotential level sets $\LevelSetfunc{\bv}{\overline{V}_N}\subset \Mfunc{\bv_0,\bv_1}{\overline{V}_N}$ is given in order to establish a link between the geometry with rescaled metric $\tilde{g}$, the geometry induced by the metric $g$, and configurational microcanonical thermodynamics.

\subsection{Geometry of Riemannian Manifolds $(\Mfunc{[\bv_0,\bv_1]}{\overline{V}_N},\tilde{g})$ and $(\LevelSetfunc{\bv}{\overline{V}_N},\tilde{g}\bigr|_{\LevelSetfunc{\bv}{\overline{V}_N}})$}

By the use of local coordinate system in $\{u_{\alpha}\}_{\alpha=0,...,(N-1)}$ we can compute the geometrical properties of the manifold $(\Mfunc{[\bv_0,\bv_1]}{\overline{V}_N},\tilde{g})$ of diffeomorphic level sets $\LevelSetfunc{\bv}{\overline{V}_N}$ endowed with the rescaled metric $\tilde{g}$ defined in \eqref{eq:rescaled_metric}.

This leads to the fact that the mean curvature of the equipotential level sets in configuration space with the rescaled
metric $\tilde{g}$ coincides with the divergence of $\overline{\xi}_N$ in configuration space \textit{with the non rescaled metric $g$}. Moreover, the curvature of the ambient manifold is related to the possibility to use some important results in differential topology (i.e. as the Chern-Lashof Theorem) which relate the global curvature integral of submanifolds immersed in space with constant scalar curvature and their topological invariants.\\
The  geometric quantities calculated in the configuration space with the rescaled metric $(\Mfunc{[\bv_0,\bv_1]}{\overline{V}_N},\tilde{g})$ are tilded and expressed in terms of the geometrical quantities calculated in $(\Mfunc{[\bv_0,\bv_1]}{\overline{V}_N},g)$ and the rescaling function $\phi_N$.

\subsubsection*{Christoffel symbols (Levi-Civita connection associated to $\tilde{g}$)}
\noindent The starting point to characterize the geometry of the configuration space, and of the regular equipotential level sets foliating it, consists in computing the Christoffel symbols associated withthe Levi-Civita connection $\tilde{\nabla}$ of the metric $\tilde{g}$.\\
Using the definition of Christoffel symbols, we obtain:
\begin{equation}
\begin{split}
&\ChristoffelTilde{0}{0}{0}=\dfrac{1}{2}\tilde{g}^{0\alpha}\left(2\partial_{0}\tilde{g}_{0\alpha}-\partial_{\alpha}\tilde{g}_{00}\right)=\dfrac{1}{2}\tilde{g}^{00}\partial_{0}\tilde{g}_{00}=
\dfrac{1}{2}g^{00}\partial_{0}g_{00}+g^{00}g_{00}\partial_{0}\phi_N=\Christoffel{0}{0}{0}-\delta_{0}^{0}\partial_{0}\phi_N=0\\
\end{split}
\end{equation}
\begin{equation}
\begin{split}
&\ChristoffelTilde{0}{i}{j}=\dfrac{1}{2}\tilde{g}^{0\alpha}\left(\partial_i
\tilde{g}_{\alpha j}+\partial_{j}\tilde{g}_{\alpha
i}-\partial_{\alpha}\tilde{g}_{ij}\right)=
-\dfrac{1}{2}\tilde{g}^{00}\partial_{0}\left(\tilde{g}_{ij}\right)=-\dfrac{1}{2}e^{2\phi_N}g^{00}\partial_{0}\left[e^{2\Xi\phi_N}g_{ij}\right]=\\
&=e^{2(\Xi+1)\phi_N}\left(\Christoffel{0}{i}{j}-\Xi g_{ij} \partial^0\phi_N\right)\\
\end{split}
\end{equation}
\begin{equation}
\begin{split}
&\ChristoffelTilde{0}{j}{0}=\dfrac{1}{2}\tilde{g}^{0\alpha}\left(\partial_j\tilde{g}_{\alpha
0}+\partial_{0}\tilde{g}_{\alpha
j}-\partial_0\tilde{g}_{0j}\right)=\dfrac{1}{2}\tilde
{g}^{00}\partial_{j}\left(\tilde{g}_{00}\right)=\dfrac{1}{2}e^{2\phi_N}
g^{00}\partial_{j}\left(e^{-2\phi_N}g_{00}\right)=\\
&=\Christoffel{0}{j}{0}-\delta^{0}_{0}\partial_{j}\phi_N=0\\
\end{split}
\end{equation}
\begin{equation}
\begin{split}
&\ChristoffelTilde{i}{0}{0}=\dfrac{1}{2}\tilde{g}^{i\alpha}\left(\partial_{0}\tilde{g}_{\alpha 0}+\partial_{0}\tilde{g}_{\alpha 0}-\partial_{\alpha}\tilde{g}_{00}\right)=\dfrac{1}{2}e^{-2\Xi\phi_N}g^{ij} \partial_{j}\left(e^{-2 \phi_N}g_{00}\right)=e^{-2(\Xi+1)\phi_N}\left[\Christoffel{i}{0}{0}+g_{00}\partial^{j}\phi_N\right]=\\
&=0\\
\end{split}
\end{equation}
\begin{equation}
\begin{split}
&\ChristoffelTilde{k}{0}{i}=\dfrac{1}{2}\tilde{g}^{k\alpha}\left(\partial_{i}\tilde{g}_{\alpha 0}+\partial_{0}\tilde{g}_{i\alpha}-\partial_{\alpha}\tilde{g}_{i}{0}\right)=\dfrac{1}{2}e^{-2\Xi\phi_N}g^{kl}\partial_{0}\left(e^{2\Xi\phi_N}g_{il}\right)=\dfrac{1}{2}g^{kl}\partial_{0}g_{li}+\Xi g^{kl}g_{li}\partial_{0}\phi_{N}=\\
&=\Christoffel{k}{0}{i}+\Xi\delta^{k}_{i}\partial_{0}\phi_N\\
\end{split}
\end{equation}
\begin{equation}
\begin{split}
\ChristoffelTilde{i}{j}{k}&=\dfrac{1}{2}\tilde{g}^{i \alpha}\left(\partial_{j}\tilde{g}_{\alpha k}+\partial_{k}\tilde{g}_{\alpha j}-\partial_{\alpha}g_{jk}\right)=\dfrac{1}{2}e^{-2\Xi\phi_N}g^{il}\left[
\partial_{j}(e^{2\Xi\phi_N}g_{li})+\partial_{k}(e^{2\Xi\phi_N}g_{lj})-\partial_{l}(e^{2\Xi\phi_N}g_{ij})\right]=\\
&=\Christoffel{i}{j}{k}+\Xi\left(\delta^{i}_{k}\partial_{j}\phi_N+\delta^{i}_{j}\partial_{k}\phi_{N}-g_{jk}\partial^{i}\phi_N\right)
\end{split}
\end{equation}

\subsubsection*{Principal curvatures, Ricci curvatures, Scalar curvature}

\noindent The expression of the sum of principal curvatures $\tilde{\tau}_{1,\tilde{g}}$ of equipotential level sets $\Sigma_{\bv}^N$ in the Riemannian manifolds $(\Mfunc{\bv_0,\bv_1}{\overline{V}_N},\tilde{g})$ is given according to the definition by
\begin{equation}
\label{eq:meancurvature_rescaledI}
\begin{split}
\tilde{\tau}_{(1,\tilde{g})}&=\widetilde{\mathrm{II}}_{ij} \tilde{g}^{ij}= \tilde{g}(\tilde{\nabla}_{\boldsymbol{\partial}_i}\tilde{\boldsymbol{\nu}}_N,\boldsymbol{\partial}_j)\tilde{g}^{ij}=\tilde{g}(\nabla_{\boldsymbol{\partial}_i}\boldsymbol{\partial}_0,\boldsymbol{\partial}_j)e^{-2\Xi\varphi_N}g^{ij}=\ChristoffelTilde{k}{0}{i}e^{2\Xi\phi_N}g_{kj}e^{2\Xi\phi_N}g^{ij}=\ChristoffelTilde{k}{0}{i}g_{kj}g^{ij}\\
&=\left(\Christoffel{k}{0}{i}+\dfrac{\delta_{k}^{i}}{N-1}\partial_{0}\phi_N\right)\delta^{k}_{i}=\Christoffel{i}{0}{i}+\dfrac{\delta^{i}_{i}}{N-1}\partial_{0}\log\left(\overline{\chi}_N\right)=\overline{\chi}_{N}\tau_{(1,g)}+\mathcal{L}_{\boldsymbol{\nu}_{N}}(\overline{\chi}_N)=\mathrm{div}_{g}\overline{\boldsymbol{\xi}}_N\\
\end{split}
\end{equation}
This gives the expected results that \textit{the sum of curvatures of equipotential level sets embedded in  configuration space with the rescaled metric $\tilde{g}$ coincides with the divergence of the vector field $\overline{\boldsymbol{\xi}}_N$ in the non rescaled configuration space.}\\
Lie derivatives of sums of principal curvatures are involved in the calculation of
the higher order derivatives of the configurational microcanonical partition function
and entropy. The formula for the first order Lie derivative of the sum of principal curvatures along
the normal field is given by
\begin{equation}
\label{eq:LieXitildetau1}
\mathcal{L}_{\tilde{\boldsymbol{\nu}}_N}(\tilde{\tau}_{1,\overline{g}})=\partial_{0}\tilde{\tau}_{1,\overline{g}}=-\tilde{\tau}_{2,\tilde{g}}-\widetilde{\mathrm{Ric}}(\tilde{\boldsymbol{\nu}}_{N},\tilde{\boldsymbol{\nu}}_N)=-\tilde{\tau}_{2,\tilde{g}}-\widetilde{\mathrm{Ric}}_{00}
\end{equation}
(its derivation is reported in \autoref{ch:DiffGeo}) where and $\widetilde{\mathrm{Ric}}$ is the Ricci tensor of ambient space with the rescaled metric $\tilde{g}$.\\
The sums of square of principal curvatures (the called "second order mean curvature")
is given by:
\begin{equation}
\begin{split}
&\tilde{\tau}_{2,\tilde{g}}=\widetilde{\mathrm{II}}_{ij}\widetilde{\mathrm{II}}_{kl}\tilde{g}^{jk}
\tilde{g}^{il}=\tilde{g}\left(\tilde{\nabla}_{\boldsymbol{\partial}_i}\boldsymbol{\partial}_{0},\boldsymbol{\partial}_j\right)\tilde{g}\left(\tilde{\nabla}_{\boldsymbol{\partial}_k}\boldsymbol{\partial}_0,\boldsymbol{\partial}_l\right)\tilde{g}^{jk}\tilde{g}^{il}=
\ChristoffelTilde{k}{i}{0}\ChristoffelTilde{i}{k}{0}=\\
&=\left(\Christoffel{k}{i}{0}+\dfrac{\delta^{i}_{k}}{(N-1)}\partial_{0}\phi_N\right)\left(\Christoffel{i}{k}{0}+\dfrac{\delta^{k}_{i}}{(N-1)}\partial_{0}\phi_N\right)=e^{2\phi_N}\tau_{2,g}+2\dfrac{e^{\phi_N}\tau_{1,g}}{(N-1)}\partial_{0}\phi_N+\dfrac{\left(\partial_{0}\phi_N\right)^2}{(N-1)} \,.
\end{split}
\end{equation}
The relevant curvature properties of the configuration space for our problem are contained  in the Ricci tensor; its contraction with the metric tensor gives the scalar Riemannian curvature of the total space that appears in many theorems and results concerning total curvature integral over immersed submanifolds.\\
From eqs.\eqref{eq:LieXitildetau1} and \eqref{eq:meancurvature_rescaledI} it is possible to derive the expression of the component $\widetilde{\mathrm{Ric}}_{00}$ in space with rescaled metric as a function of the same component for the Ricci tensor in space with metric $g$:
\begin{equation}
\begin{split}
&\widetilde{\mathrm{Ric}}_{00}=\widetilde{\mathrm{Ric}}(\widetilde{\boldsymbol{\nu}}_N,\widetilde{\boldsymbol{\nu}}_N)=-\partial_0\left(\tilde{\tau}_{1,g}\right)-\tilde{\tau}_{2,g}=\mathrm{Ric}_{00}-\dfrac{N+1}{N-1}\Christoffel{i}{i}{0}\partial_0\phi_N-\Xi\partial_{0}\phi_N\partial_{0}\phi_N-\partial_{0}^2\phi_N
\end{split}
\end{equation}
The Ricci tensor $\widetilde{\left(\mathrm{Ric}_{\Sigma}\right)}_{ij}$ restricted over the potential level sets which transforms under conformal changes as follows (see \cite{besse2007einstein}) 
\begin{equation}
\label{eq:RicciSigma}
\begin{split}
&\widetilde{\left(\mathrm{Ric}_{\Sigma}\right)}_{ij}=\partial_k \ChristoffelTilde{k}{i}{j}-\partial_{j}\ChristoffelTilde{k}{k}{i}+\ChristoffelTilde{k}{k}{l}\ChristoffelTilde{l}{i}{j}-\ChristoffelTilde{k}{j}{l}\ChristoffelTilde{l}{k}{i}=\left(\mathrm{Ric}_{\Sigma}\right)_{ij}-\dfrac{N-3}{N-1}\left[\nabla_{\boldsymbol{\partial}_i}\partial_j\phi_N-\dfrac{\left(\partial_i\phi_N\right)\left(\partial_j \phi_N\right)}{N-1}\right]+\\
&+\dfrac{(N-1)\Delta_{\Sigma} \phi_N-(N-2)\|\mathbf{grad}_{g_{\Sigma}} \phi_N\|_{g_{\Sigma}}^2}{(N-1)^2}g_{ij}=\left(\mathrm{Ric}_{\Sigma}\right)_{ij}-F^{(1)}_{ij}+F^{(2)}g_{ij}
\end{split}
\end{equation}
being $\Delta_{\Sigma}$ the Laplace-Beltrami operator restricted on the regular potential level set
\begin{equation}
\Delta_{\Sigma} f=g^{ij}\Christoffel{k}{i}{j}\partial_{k}f-\partial_{i}\partial^{i}f\,.
\end{equation}
The components of the Ricci tensor along tangent direction to level sets $\LevelSetfunc{\bv}{\overline{V}_N}$ are given by
\begin{equation}
\begin{split}
&\widetilde{\mathrm{Ric}}_{ij}=\partial_0 \ChristoffelTilde{0}{j}{i}-\partial_j \ChristoffelTilde{0}{0}{i}+\ChristoffelTilde{0}{0}{0}\ChristoffelTilde{0}{i}{j}+\ChristoffelTilde{0}{0}{k}\ChristoffelTilde{k}{i}{j}+\ChristoffelTilde{k}{k}{0}\ChristoffelTilde{0}{j}{i}-\ChristoffelTilde{0}{j}{0}\ChristoffelTilde{0}{i}{0}-\ChristoffelTilde{0}{j}{k}\ChristoffelTilde{k}{0}{i}-\ChristoffelTilde{k}{j}{0}\ChristoffelTilde{0}{k}{i}+\widetilde{\left(\mathrm{Ric}_{\Sigma}\right)}_{ij}=\\
&=\mathrm{Ric}_{ij}+\left(e^{2(\Xi+1)\phi_N}-1\right)\left[\partial_0\Christoffel{0}{i}{j}+\Christoffel{0}{0}{0}\Christoffel{0}{i}{j}+\Christoffel{0}{i}{j}\Christoffel{k}{0}{k}-\left(\Christoffel{0}{j}{k}\Christoffel{k}{0}{i}+\Christoffel{k}{j}{0}\Christoffel{0}{i}{k}\right)\right]+\\
&+2e^{2(\Xi+1)\phi_N}\partial_0\phi_N\Christoffel{0}{i}{j}+\partial_{j}\partial_{i}\phi_N-\Christoffel{k}{i}{j}\partial_{k}\phi_N+\partial_j\phi_N\partial_i\phi_N+\\
&-g_{ij}e^{2(\Xi+1)\phi_N}\Xi\left(3\partial^{0}\phi_N\partial_{0}\phi_N+\partial_{0}\partial^{0}\phi_N+\Christoffel{k}{0}{k}\partial^{0}\phi_N\right)-F^{(1)}_{ij}+F^{(2)}g_{ij}\\
\end{split}\,.
\end{equation}
The other components of Ricci tensor in rescaled metric read:
\begin{equation}
\begin{split}
&\widetilde{\mathrm{Ric}}_{0i}=\partial_{\alpha}\ChristoffelTilde{\alpha}{i}{0}-\partial_{i}\Christoffel{\alpha}{\alpha}{0}+\Christoffel{\alpha}{\alpha}{\beta}\Christoffel{\beta}{i}{j}-\Christoffel{\alpha}{\beta}{0}\Christoffel{\beta}{\alpha}{0}=\mathrm{Ric}_{i0}-\dfrac{N-2}{N-1}\partial_{0}\partial_{i}\phi_N-\Christoffel{0}{0}{0}\partial_{i}\phi_N+\\
&-\dfrac{N}{N-1}\Christoffel{j}{j}{0}\partial_{i}\phi_N+\dfrac{N-2}{N-1}\Christoffel{k}{i}{0}\partial_{k}\phi_N+\Christoffel{0}{i}{0}\partial_{0}\phi_N-g_{00}\Christoffel{0}{i}{j}\partial^{j}\phi_N+g_{ik}\partial^{j}\Christoffel{k}{j}{0}
\end{split}
\end{equation}

It follows that the scalar curvature of the equipotential level sets is given
by 
\begin{equation}
\mathcal{R}_{\Sigma}=\tilde{g}^{ij}\widetilde{\left(\mathrm{Ric}_{\Sigma}\right)_{ij}}=e^{-2\Xi \phi_N}\left(\mathcal{R}_{\Sigma}+2\dfrac{(N-2)}{(N-1)}\Delta\phi_N-\dfrac{(N-3)(N-2)}{(N-1)^2}\|\mathbf{grad}_g\phi_N\|^2_g\right)
\end{equation}
while the scalar curvature of the ambient space $\tilde{\mathcal{R}}$ in rescaled metric reads
\begin{equation}
\begin{split}
&\widetilde{\mathcal{R}}=\widetilde{g}^{\alpha \beta}\widetilde{\mathrm{Ric}}_{\alpha\beta}=\widetilde{g}^{00}\widetilde{\mathrm{Ric}}_{00}+\widetilde{g}^{ij}\widetilde{\mathrm{Ric}}_{ij}=e^{2\phi_N}g^{00}\mathrm{Ric}_{00}+\\
&+e^{-2\Xi\phi_N}g^{ij}\mathrm{Ric}_{ij}-e^{2\phi_N}\left[\dfrac{2N}{N-1}\Christoffel{i}{i}{0}\phi_N+\dfrac{3N}{N-1}\partial^{0}\phi_N\partial_{0}\phi_N-2\partial_{0}\partial^{0}\phi_N\right]+\\
&+\left(e^{2\phi_N}-e^{2\Xi\phi_N}\right)g^{ij}\left[\partial_0\Christoffel{0}{i}{j}+\Christoffel{0}{0}{0}\Christoffel{0}{i}{j}+\Christoffel{0}{i}{j}\Christoffel{k}{0}{k}-\left(\Christoffel{0}{j}{k}\Christoffel{k}{0}{i}+\Christoffel{k}{j}{0}\Christoffel{0}{i}{k}\right)\right]+2e^{2\phi_N}\partial_0\Christoffel{0}{i}{j}g^{ij}+\\
&+\dfrac{3N-5}{N-1}e^{-2\phi_N\Xi}\left(\Delta_{\Sigma}\phi_N+\|\mathbf{grad}_{\Sigma}\phi_N\|^2_{g_{\Sigma}}\right)\\
\end{split}
\end{equation}
\subsection{Geometrical interpretation of the configurational microcanonical statistical mechanics}
The metric rescaling introduced above allows to give a \textit{pure geometrical interpretation of configurational microcanonical entropy and its derivatives}.
If we apply the rule proved in eq.\eqref{eq:Derivation_in_intII} to pass the derivatives of the control parameter $\bv$ into the integral of\eqref{eq:rescMetr_Omega} we obtain:
\begin{equation}
 \begin{split}
 \dfrac{\D \Omega_N}{\D \bv}&=\intSigmaV{\bv}{\overline{V}_N}\,\mathcal{L}_{\bv}(\D
 \sigma_{\LevelSetfunc{\bv}{\overline{V}_N},\tilde{g}})=\intSigmaV{\bv}{\overline{V}_N}\,\tau_{1,\tilde{g}}\,\,\D
 \sigma_{\LevelSetfunc{\bv}{\overline{V}_N},\tilde{g}}\\
 \dfrac{\D^2 \Omega_N}{\D
 \bv^2}&=\intSigmaV{\bv}{\overline{V}_N}\,\mathcal{L}_{\bv}(\tau_{1,\tilde{g}}\,\D
 \sigma_{\LevelSetfunc{\bv}{\overline{V}_N},\tilde{g}})=\intSigmaV{\bv}{\overline{V}_N}\,\,\left[\tau_{1,\tilde{g}}^2+\mathcal{L}_{\boldsymbol{\nu}_N}(\tau_{1,\tilde{g}})\right]\,\,\D
 \sigma_{\LevelSetfunc{\bv}{\overline{V}_N},\tilde{g}}\\
  \dfrac{\D^3 \Omega_N}{\D
 \bv^3}&=\intSigmaV{\bv}{\overline{V}_N}\,\left[\tau_{1,\tilde{g}}^3+3\tau_{1,\tilde{g}}\mathcal{L}_{\boldsymbol{\nu}_N}(\tau_{1,\tilde{g}})+\mathcal{L}_{\boldsymbol{\nu}_N}\left(\mathcal{L}_{\boldsymbol{\nu}_N}(\tau_{1,\tilde{g}})\right)\right]\,\,\D
 \sigma_{\LevelSetfunc{\bv}{\overline{V}_N},\tilde{g}}\\
  \dfrac{\D^4 \Omega_N}{\D
 \bv^4}&=\intSigmaV{\bv}{\overline{V}_N}\Bigr[\tau_{1,\tilde{g}}^4+6
 \tau_{1,\tilde{g}}^2\mathcal{L}_{\boldsymbol{\nu}_N}(\tau_{1,\tilde{g}})+4\tau_{1,\tilde{g}}\mathcal{L}_{\boldsymbol{\nu}_N}(\mathcal{L}_{\boldsymbol{\nu}_N}(\tau_{1,\tilde{g}}))+\\
&+3\tau_{1,\tilde{g}}\left(\mathcal{L}_{\boldsymbol{\nu}}(\tau_{1,\tilde{g}})\right)^2+
\mathcal{L}_{\boldsymbol{\nu}_N}(\mathcal{L}_{\boldsymbol{\nu}_N}(\mathcal{L}_{\boldsymbol{\nu}_N}(\tau_{1,\tilde{g}})))\Bigr]
\D\sigma_{\LevelSetfunc{\bv}{\overline{V}_N},\tilde{g}}
 \end{split}
 \end{equation}
and consequently, using eqs.\eqref{def:defintition_statQuant} and \eqref{eq:microcanEntropy_Derivative_Xi}, and the definition of mean curvature in eq.\eqref{eq:def_Meancurvature} we obtain for the derivatives of specific configurational microcanonical entropy:
\begin{align}
\label{eq:Geometrize_EntropyDerivatives}
&\dfrac{\D\overline{S}_N}{\D \bv}(\bv)=\dfrac{1}{N}\left\langle \tau_{1,\tilde{g}}
\right\rangle_{\bv,\tilde{\sigma}}=\dfrac{(N-1)}{N}\left\langle
h_{\tilde{g}} \right\rangle_{\bv,\tilde{\sigma}}\\
&\dfrac{\D^2\overline{S}_N}{\D \bv^2}(\bv)=\dfrac{1}{N}\left[\left\langle \tau_{1,\tilde{g}}^2 \right\rangle_{\bv,\tilde{\sigma}}+\left\langle
 \mathcal{L}_{\nu_N}(\tau_{1,\tilde{g}})\right\rangle_{\bv,\tilde{\sigma}}^2-\left\langle
\tau_{1,\tilde{g}}\right\rangle_{N\bv ,\tilde{\sigma}}^2\right]=\\
&=\dfrac{(N-1)}{N}\left[(N-1)\mathrm{Var}_{N\bv,\tilde{\sigma}}(h_{\tilde{g}})+\left\langle\mathcal{L}_{\nu_N}
(h_{\tilde{g}}) \right\rangle_{N\bv,\tilde{\sigma}}\right]\\
&\dfrac{\D^3\overline{S}_N}{\D \bv^3}=\dfrac{1}{N}\Bigr[\left \langle \tau_{1,\tilde{g}}^3 \right\rangle_{\bv,\tilde{\sigma}}+3\left \langle \tau_{1,\tilde{g}}\mathcal{L}_{\nu_N}(\tau_{1,\tilde{g}}) \right\rangle_{\bv,\tilde{\sigma}}+ 
\left\langle \mathcal{L}_{\nu_N}^{(ii)}(\tau_{1,\tilde{g}}) \right\rangle_{\bv,\tilde{\sigma}}+\\
&+3\left \langle \tau_{1,\tilde{g}} \right\rangle_{\bv,\tilde{\sigma}}\left \langle\mathcal{L}_{\nu_N}(\tau_{1,\tilde{g}}) \right\rangle_{\bv,\tilde{\sigma}}-3 \left \langle \tau_{1,\tilde{g}}\right\rangle_{\bv,\tilde{\sigma}}^2\left \langle \tau_{1,\tilde{g}}\right\rangle_{\bv,\tilde{\sigma}}+2\left \langle \tau_{1,\tilde{g}}\right\rangle_{\bv,\tilde{\sigma}}^3\Bigr]=\\
&=\dfrac{(N-1)}{N}\left[(N-1)^2\mathrm{Cumul}^{(3)}_{\bv,\sigma}(h_{1,\tilde{g}})+(N-1)\mathrm{Cov}_{\bv,\sigma}(h_{1,\tilde{g}};\mathcal{L}_{\nu_N}(h_{1,\tilde{g}}))+\left\langle \mathcal{L}_{\nu_N}^{(ii)}(h_{1,\tilde{g}}) \right\rangle_{\bv,\tilde{\sigma}}\right]\\
&\dfrac{\D^4 \overline{S}_N }{\D
\overline{v}^4}(\bv)=\dfrac{1}{N}\Biggr[\mathrm{Cuml}^{(4)}_{\bv,\mu}(\tau_{1,\tilde{g}})+
6\mathrm{Cov}_{\bv,\mu}\left(\tau_{1,\tilde{g}};\tau_{1,\tilde{g}}\mathcal{L}_{\nu_N}(\tau_{1,\tilde{g}})\right)+
3\mathrm{Var}_{\bv,\mu}\left(\mathcal{L}_{\nu_N}(\tau_{1,\tilde{g}})\right)+\\
&+4\mathrm{Cov}_{\bv,\mu}\left(\tau_{1,\tilde{g}};\mathcal{L}_{\nu_N}^{(ii)}(\tau_{1,\tilde{g}})\right)
+12\mathrm{Cov}_{\bv,\mu}\left(\tau_{1,\tilde{g}}^2;\mathcal{L}_{\nu_N}(\tau_{1,\tilde{g}})\right)+
\left\langle\mathcal{L}_{\nu_N}^{(iii)}\left(\tau_{1,\tilde{g}}\right)\right\rangle_{\bv,\mu}
\Biggr]=\\
&=\dfrac{(N-1)}{N}\Biggr[(N-1)^3\mathrm{Cuml}^{(4)}_{\bv,\mu}(h_{1,\tilde{g}})+
6(N-1)^2\mathrm{Cov}_{\bv,\mu}\left(h_{1,\tilde{g}};h_{1,\tilde{g}}\mathcal{L}_{\nu_N}(h_{1,\tilde{g}})\right)+\\
&+12(N-1)^2\mathrm{Cov}_{\bv,\mu}\left(h_{1,\tilde{g}}^2;\mathcal{L}_{\nu_N}(h_{1,\tilde{g}})\right)+3(N-1)\mathrm{Var}_{\bv,\mu}\left(\mathcal{L}_{\nu_N}(h_{1,\tilde{g}})\right)+\\
&+4(N-1)\mathrm{Cov}_{\bv,\mu}\left(h_{1,\tilde{g}};\mathcal{L}_{\nu_N}^{(ii)}(h_{1,\tilde{g}})\right)+\left\langle\mathcal{L}_{\nu_N}^{(iii)}\left(h_{1,\tilde{g}}\right)\right\rangle_{\bv,\mu}
\Biggr]\\
\end{align}
where the statistical quantities are calculated over the level sets
$\LevelSetfunc{\bv}{\overline{V}_N}$ using the induced Riemannian area form
$\D\sigma_{\LevelSetfunc{\bv}{\overline{V}_N},\tilde{g}}$ that coincides with the
microcanonical area measure.\\
It has to be stressed that in the Riemannian formulation proposed for the
configurational microcanonical ensemble \textbf{the inverse of microcanonical temperature
$\beta(\bv)=T^{-1}(\bv)=(\partial \overline{S}_N/\partial \bv)$ coincides for large 
$N$ with the average of the mean curvature of the associated equipotential level set
$\LevelSetfunc{\bv}{\overline{V}_N}$ giving a quite simple geometrical interpretation of 
this basic statistical mechanics observable}.\\
More in general, the geometrical interpretation of statistical mechanics in configurational 
microcanonical ensemble can help the research of other signatures at finite $N$, with 
respect to topological changes, that signal the presence of phase transitions. In i 
particular we remark that an interesting possible starting point for further investigations
consists in the formalization in geometrical terms of a condition that can allow to 
prevent the second order derivative of entropy to be non-negative.
In fact, according to the theory of microcanonical statistical analysis a signature of phase transition in the microcanonical ensemble
is the non-concavity of the entropy, i.e. $\partial_{\bar{E}}^2\overline{S}_N\geq 0$ or $\lim_{N\rightarrow+\infty}\partial_{\bar{E}}^2\overline{S}_N(\bar{E}_c)=0$.\\
Assuming that the same signature of phase transitions is observed also in configuration space, the condition that has to be imposed to prevent phase transitions in thermodynamic limit is given by
\begin{equation}
\label{eq:Cond_Var_LieXih}
\mathrm{Var}_{\bv,\overline{\sigma}}\tilde{h}_{1,\tilde{g}}\leq B<\left\langle \mathcal{L}_{\tilde{\boldsymbol{\nu}}_N}\tilde{h}_{1,\tilde{g}}\right\rangle \qquad \forall N\in\mathbb{N}\,\,.
\end{equation}
In this framework, the problem of phase transitions could be formulated in terms of suitable condition that has to be imposed to the mean curvature field $\tilde{h}_{1,\tilde{g}}$ in order to satisfy the condition in eq.\eqref{eq:Cond_Var_LieXih} and that can be possibly read as a geometrical characterization of a topological asymptotic change.\\
The most difficult issue to overcome in this scheme of refinement of the Necessity Theorem is constituted by the derivation of an upper bound of the variance of the mean curvature over a level set. This aspect is still an open problem; nevertheless some possible strategies to derive upper bounds of the variance have been investigated. A promising starting point to attain this aim would be the Poincaré Inequality \cite{ledoux2005concentration}
\begin{equation}
\lambda_1\mathrm{Var}_{\bv,\overline{\sigma}} f\leq\left\langle \|\mathbf{grad}_g f\|^2\right\rangle_{\bv,\overline{\sigma}} 
\end{equation}
where $\lambda_1$ is the first non trivial eigenvalues of the Laplace-Beltrami operator.
Lower bounds on $\lambda_1$ for compact riemannian manifolds $(M,g)$ with non-negative Ricci curvature have been obtained by Li-Yau and Zhong-Yang (see \cite{li1993lecture} and reference therein)
\begin{equation}
\lambda_{1,g} (M)\geq\dfrac{\pi^2}{\mathrm{diam}_g^2(M)}
\end{equation}
where $\mathrm{diam}_g(M)$ is the diameter of the manifold.
 Another important results is the
Lichnerowicz theorem:
\begin{theorem}[Lichnerowicz theorem]
et M be an $N$-dimensional compact manifold
without boundary. Suppose that the Ricci curvature of M is bounded from below
by
\begin{equation}
\mathrm{Ric}_{ij}\geq (N-1)K
\end{equation}
for some constant $K>0$, then the first nonzero eigenvalue of the Laplacian on M must
satisfy
\begin{equation}
\lambda_{1,g}(M) \geq NK
\end{equation}
Moreover, equality holds if and only if M is isometric to a standard sphere of radius
$K^{-1/2}$.
\end{theorem}
Despite of the fact that these results would seem to suggest the possibility to easily construct non tautological geometrical conditions on the mean curvature $\tilde{h}_{1,\tilde{g}}$, the condition of non-negative Ricci curvature $\widetilde{\left(\mathrm{Ric}\right)}_{ij}$ seems to not be easily verifiable \textit{a priori} from eq.\eqref{eq:RicciSigma} nor \textit{a posteriori} from the results of numerical simulations on equipotential level sets as it is a pointwise condition.

\section{Conclusions}
In the present work we have discussed how it is possible to encode
the full information on the thermodynamic behavior of a 
given system into a riemannian geometric structure defined in 
subspaces of configuration space where equipotential level sets define a regular foliation.\\
This result could in principle pave the way for new approaches to 
study how phase transition in microcanonical ensemble are related 
with the geometry and topolpogy of equipotential level sets. Some of
these new developments are presented in a recent paper 
\cite{di2022geometric} where it is studied in details the interplay
between the riemannian curvatures of the configuration space and the
rate expansion of the configurational microcanonical volume $W_{N}(\bar{v})$.\\

\appendix
\label{ch:DiffGeo}

\section{Brief review of Riemannian Geometry}

In this appendix we review some concepts concerning the Riemannian Geometry
that occurring in the main part of the manuscript. The reader is supposed to
be acquainted with very basic notions of differential geometry that we briefly
sketch here.

\subsection{The concept of differentiable manifold}
\noindent One of the main purpose of the {\bf Differential Geometry} is to extend the applicability of the differential calculus, usually performed on open set of the vector (linear) space $\R^n$, to more general sets. Basically, the definitions and the principal features of the differential calculus are local properties; i.e., they depend only on their behaviour in an arbitrarily small neighbourhood of a point. So, if a set $\cal M$ is locally as an open set of $\R^n$, we are able to introduce a differential calculus on $\cal M$. 

An $n$-Smooth Manifold is an abstract set\footnote{Here we introduce the concept of the manifold thinking of it as living into no linear space $\mathbb{R}^n$.} such that a small region around each point of it is done as an open set\footnote{This requirement implies that the manifold has the same topology of $\R ^n$, at least locally.} of $\mathbb{R}^n$ equipped with an additional smooth structure we are going to introduce.
A set $\cal M$ which is locally like the linear space $\mathbb{R}^n$ is called a \textit{topological manifold} of dimension $n$; here, an inhabitant of $\cal M$, living a neighbourhood of a given point $p$, needs exactly $n$ dimensions to describe the surrounding reality. Roughly speaking, the map $\varphi(p)=(x^1(p),\ldots,x^n(p))$ realizes those dimensions on $\R^n$ and the maps $x^1(p),\ldots,x^n(p)$ are called \textit{local coordinates} of $\cal M$ in $p$. More formally, given a set $\cal M$ and $p\in \mathcal{M}$, a chart $(U,\varphi)$ at $p$ is a bijective map $\varphi:U\rightarrow V$, where $U\subset\mathcal{M}$ and $V$ is an open set of $\R^n$. Actually, a chart provides on $U\subset\mathcal{M}$ the coordinates of $\R^n$. In this way, a chart allows one to describe each point of $U$ with a $n$-tuple of real numbers. Thus, $U$ is the part of $\cal M$ that is essentially like $\R ^n$. Roughly speaking, the chart $(U,\varphi)$ provides a geographic map to describe $\cal M$, at least in a small part of it. It is then clear that to describe the whole set $\cal M$ we need a collection of charts covering all $\cal M$. 

\begin{figure}[h!] \centering
\includegraphics[scale=0.2]{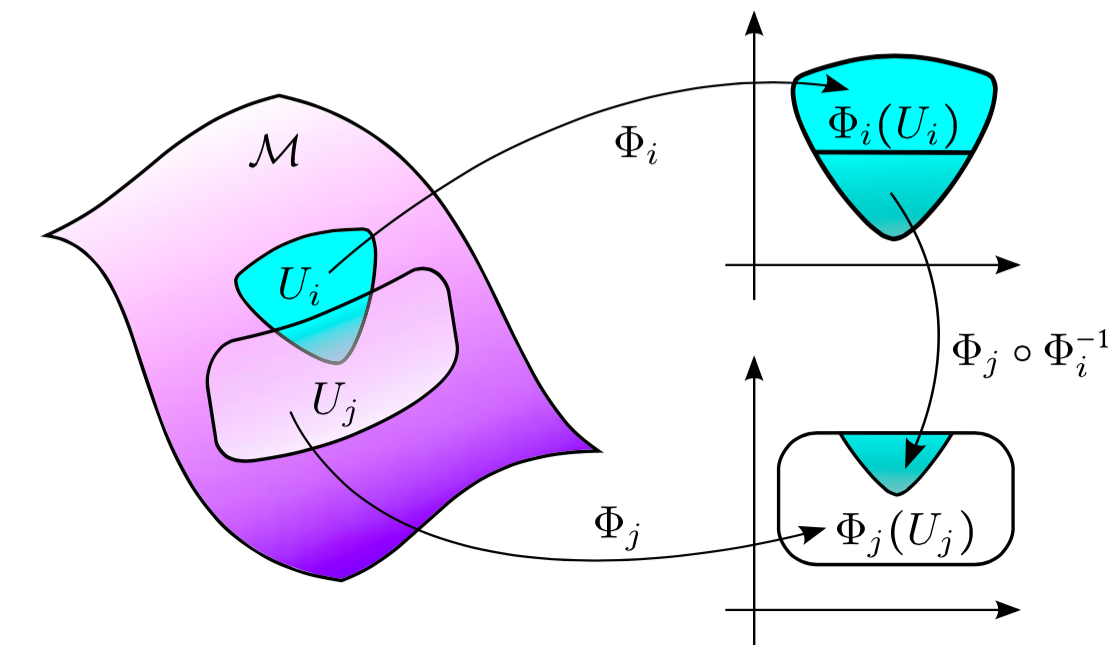}
\caption[10 pt]{\textbf{The transition map}: the sets $U_i,U_j\subset\mathcal{M}$ have $\Phi_i(U_i)$ and $\Phi_j(U_j)$ as realization into the linear space $\R ^2$. The overlap $U_i\cap U_j$ should have the same realization with both $\Phi_i$ and $\Phi_j$, as it turns out in the turquoise triangle.}
\label{cambio}
\end{figure}

A problem arises when a region of $\cal M$ is described by different charts: this chance is drawn in Fig. \ref{cambio}. Consider, for example, two inhabitants of $\cal M$ living one in a neighbourhood of a point $p\in\mathcal{M}$ and the other in a neighbourhood of a point $q\in\mathcal{M}$. They could display the respective surroundings with two different charts, say $(U_1,\varphi_1)$, $(U_2,\varphi_2)$ and so with two different sets of \textit{local coordinates}. Moreover, in case of overlap, i.e. in the region $U=U_1\cap U_2\neq \emptyset$, the charts should be (in a suitable sense) equivalent. The equivalence is provided by the compatibility condition: $ \varphi_1(U)$ and $\varphi_2(U)$ are open sets of $\R ^n$ and $\varphi_2\circ \varphi_1^{-1}:\varphi_1(U)\rightarrow\varphi_2(U)$ is a bijective, smooth map with inverse smooth again\footnote{This is properly the definition of a \textit{diffeomorphism}.}. The map $\varphi_2\circ \varphi_1^{-1}$ is called change of coordinates or \textit{transition map}. This map allows $U$ to inherit the structure of the linear space $\R ^n$.

Finally, a natural covering of the set $\cal M$ is provided by a collection of compatible (each one with each other) charts. In this way all regions of $\cal M$ can be described by equivalent local coordinates. A collection of compatible charts covering $\cal M$ is called \textit{atlas}. Moreover, two atlas are compatible if their union is yet an atlas of $\cal M$. A \textit{differentiable structure} is the maximal (with respect the inclusion) \textit{atlas} $\cal M$. Thus, a smooth manifold is the pair $(\mathcal{M},\mathcal{A})$ where $\cal A$ is a differentiable structure.
Trivially, $\R ^n$ is a smooth manifold with one chart and local coordinates given by the natural ones. 

\subsection{Tangent and cotangent space}
\begin{figure}[h!] \centering
\includegraphics[scale=0.2]{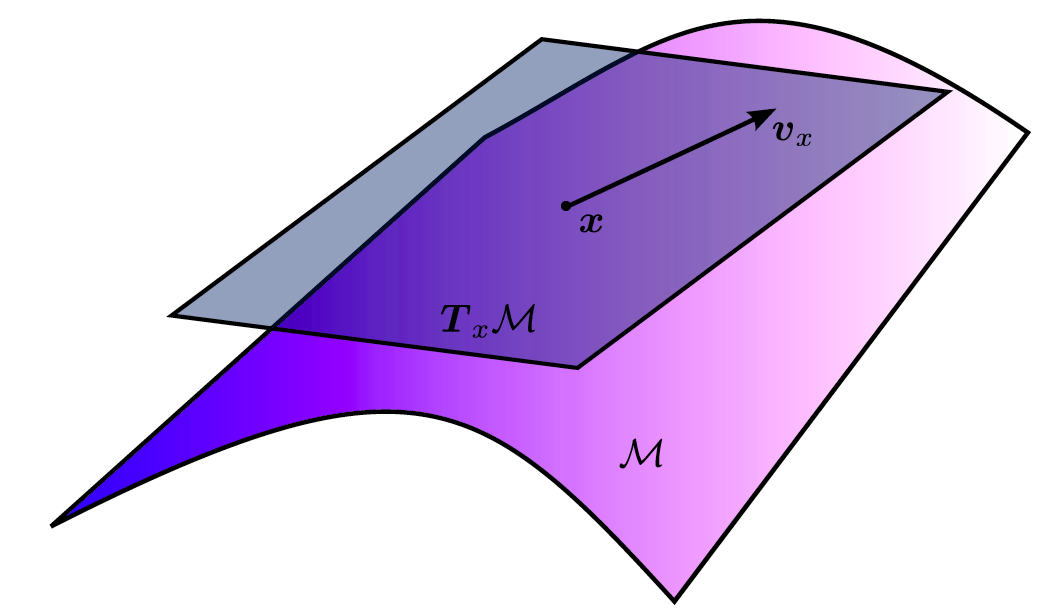}
\caption[10 pt]{\textbf{The tangent space of $\cal M$ at $x$}: the vector $v_x$ is the tangent vector of a suitable curve $\sigma(t)$ living in a small region of $\cal M$, such that $\sigma(0)=x$.}
\label{tangent}
\end{figure}

Other extra structures on the manifold $\cal M$ are possible. The most evident example of manifold with an extra structure is the linear space $\R ^n$: it is also a vector space. The canonical scalar product allows to introduce a metric structure on $\R ^n$: we can measure the length of the tangent vectors, the length of the curves and the distance between two different points. Now we want introduce a metric structure also on $\cal M$. Doing that we obviously need a structure of vector space associated with $\cal M$. We know that a small region is like  an open set\footnote{This means, the small region of $\cal M$ has the same topology of $\R ^n$.} $U\subset \R ^n$, but as matter of fact we can not add two different points $p$ and $q$ in $\cal M$. The associated structure of vector space is given by the definition of the {\bf tangent space} of $\cal M$ at a given point $p\in\mathcal{M}$. Thus, at any point $p$ we have a vector structure and so a scalar product. Then, roughly speaking, we can consider a metric structure on $\cal M$ as the union of all scalar product on any tangent space as $p$ varies on $\cal M$. Let us see first what is a tangent vector of $\cal M$ at $p$ and then which is the structure of the tangent space. Consider another mathematical object with one dimension, the \textit{curve}, different from the \textit{line}\footnote{The number eight drawn in a $\R ^2$ is a \textit{curve}, but not a \textit{line}. In fact, any small region around the auto-intersection point is a cross, but not a straight line, as request from the definition of the \textit{line}.}. More formally, a smooth \textit{curve} is a smooth map $\sigma: I\rightarrow\R^n $, where $I\subset\R $ is an interval. Given $p\in\mathcal{M}$, we can realize a small region around $p$ as an open set $U\subset\R ^n$ and perform the derivative of any differentiable function defined on it. A \textit{tangent vector} of $\cal M$ at $p$ is the one $\sigma^\prime(0)$, where $\sigma:(-\delta,\delta)\rightarrow \mathcal{M}$ is a curve inside the small region around $p$ and such that $\sigma(0)=p$: see Fig. \ref{tangent} for a major understanding of the concept. The collection of all tangent vectors of $\cal M$ at $p$ is the tangent space $T_p{\cal M}$ of $\cal M$ at $p$.
In general, given an $n$-dimensional smooth manifold $\cal M$ and a point $p\in\mathcal{M}$, the tangent space $T_p\mathcal{M}$ of $\cal M$ at $p$ is a vector space which could be identified with the set of the partial derivatives. So, let $(U,\varphi)$ be the chart providing the realization of the small region around $p$ as an open set of $\R ^n$, a basis of $T_p\mathcal{M}$ is given by $\mathcal{B}=\{\boldsymbol{\partial}_1,\ldots,\boldsymbol{\partial}_n\}$, where $\boldsymbol{\partial}_i=\frac{\partial}{\partial x^i}$ and $\{x^i\}_{i=1}^n$ is the set of local coordinates.\\
The vector space structure $T_p\mathcal{M}$ naturally allows to define a dual vector space, the so called \textbf{cotangent space} $T_p^{*}\mathcal{M}$, such that every; namely if $\theta\in T_p^{*}\mathcal{M}$ and $\boldsymbol{X}\in T_p\mathcal{M}$ then $\theta_p(\boldsymbol{X})\in\R$. In particular, if $\mathcal{B}^*=\{dx_p^1,\ldots,dx_p^n\}$ is a basis of $T_p^*\mathcal{M}$ then $dx_p^i(\partial_j)=\delta_{ij}$\footnote{$\mathcal{B}^*$ is the dual basis of $\mathcal{B}$ and the Kronecher symbol $\delta_{ij}$ is defined as $\delta_{ij}=1$ if $i=j$, otherwise it is zero.}\\
This two vector space allow to construct a \textit{tensor space} of rank $(r,s)$, a multilinear map of the form
\begin{equation}
\mathcal{T}^{(r)}_{(s),p}=\otimes_{r} T_p\mathcal{M}\otimes_{s}T^{*}_{p}\mathcal{M}
\end{equation}
an element $t\in \mathcal{T}^{(r)}_{(s),p}$ that in local components read
\begin{equation}
t=t^{i_1\dots i_r}_{j_1\ldots j_s}\boldsymbol{\partial}_{i_1}\otimes\boldsymbol{\partial}_{i_r}\otimes \D x_p^{j_1}\otimes \D x_p^{j_s}\,.
\end{equation}
Fundamental operation on the tensor and their multilinear algebra are
\begin{itemize}
\item the \textit{linear combination of two tensor} that in coordinates reads
$\left(\alpha x+\beta y\right)^{i_1 \ldots i_r}_{j_1\ldots j_s}=\alpha x^{i_1 \ldots i_r}_{j_1\ldots j_s}+\beta y^{i_1 \ldots i_r}_{j_1\ldots j_s}$ where $\alpha,\beta\in\mathbb{R}$;

\item let $1\leq\lambda\leq r$ and $1\leq \mu \leq s$ and $x\in \mathcal{T}^{(r)}_{(s),p}$  then its \textit{tensor contraction} $C_{\lambda\mu}(x)\in \mathcal{T}^{(r-1)}_{(s-1),p}$ reads in coordinate $(C_{\lambda \mu}(x))^{i_1 \ldots i_r}_{j_1\ldots j_s}=x^{i_1 \ldots i_{\lambda-1} k i_{\lambda+1}\ldots i_r}_{j_1\ldots j_{\mu-1} k j_{\mu+1}\ldots j_s}$
\item the \textit{tensor product} of two tensor $x\in x\in \mathcal{T}^{(r_1)}_{(s_1),p}$
and $x\in \mathcal{T}^{(r_2)}_{(s_2),p}$ whose action in coordinates reads
\begin{equation}
\left(x\otimes y\right)^{i_1 \ldots i_{r_1+r_2}}_{j_1\ldots j_{s_1+s_2}}=^{i_1 \ldots i_{r_1}}_{j_1\ldots j_{s_1}}\cdot y^{i_{r_1+1} \ldots i_{r_1+r_2}}_{j_{s_1+1}\ldots j_{s_1+s_2}}
\end{equation}
this operation satisfies associative and distributive laws;
\item the \textit{index permutation} $\sigma$ that can be applied to tensors of rank $(r,0)$ (multivector) or $(0,s)$ (multiform); for instance, in the latest case
we have that $\left(\sigma(x)\right)_{j_1\ldots j_s}=x_{j_{\sigma{1}}\ldots j_{\sigma{s}}}$.
If we define the sign of a permutation as 
\begin{equation}
\mathrm{sgn}\sigma=
\begin{cases}
+1 \qquad \text{even permutations}\\
-1 \qquad \text{odd permutations}\\
\end{cases}
\end{equation}
we can introduce the \textit{symmetrazing map} $S_{s}$ and the \textit{alternating} map $A_s$ of a multiform $x\in \mathcal{T}^{(0)}_{(s),p}$
\begin{align}
A_s(x)=\dfrac{1}{s!}\sum_{\sigma\in\mathcal{S}(s)}\,\,\mathrm{sgn}\sigma\cdot\sigma{x}
S_s(x)=\dfrac{1}{s!}\sum_{\sigma\in\mathcal{S}(s)}\,\,\sigma{x}\,\,.
\end{align}
\end{itemize} 
The alternating map allows to define an \textit{exterior (wedge) product} of an $s$-form
$\eta$ and an $k$-form $\rho$ as
\begin{equation}
\eta \wedge \rho=A_{s+k}(\eta\otimes\rho)
\end{equation}
with associative law, distributive law respect to sum and anticommutative law, i.e.
$\rho \wedge \theta=(-1)^{ks} \theta \wedge \rho$. The local space of $k$-forms will be indicated in what follows with $\Lambda^{k}_{p}$.

\section{Tensor fields, derivations, connections and curvatures}

\begin{figure}[h!] \centering
\includegraphics[scale=0.2]{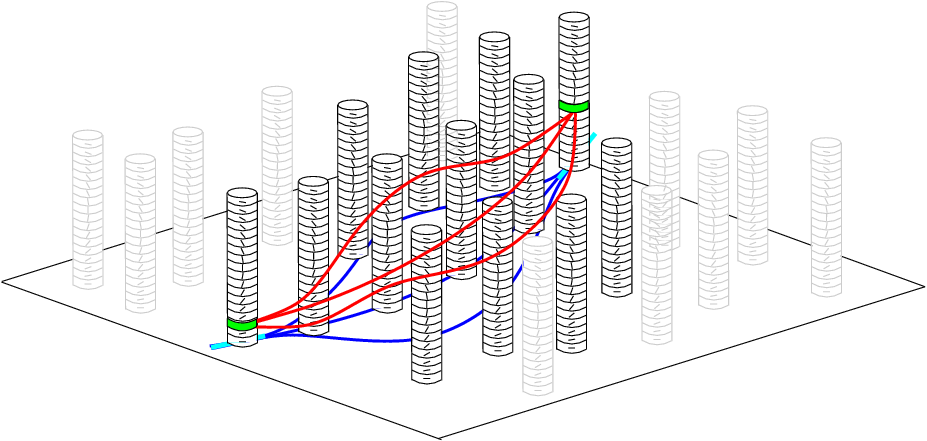}
\caption[10 pt]{\textbf{The tangent bundle}: the columns represent the tangent spaces, called, in general, fibers of bundle. The blue lines are curves on the manifold, while the red lines are indicated as the sections of the tangent bundle upon the blue curve.}
\label{tangentbundle}
\end{figure}

The vector (or tensor) space introduced in the previous section are defined point-wise; the 
next step to enrich the structure of a differential manifolds consists in extend such structures to the whole manifold. 
A very simple way to realise this consists in taking the disjoint union $T\mathcal{M}$ of all $T_p\mathcal{M}$ with respect $p\in\mathcal{M}$ : this is called the \textit{tangent bundle}. Analogously the disjoint union $T^*\mathcal{M}$ of all $T_p^*\mathcal{M}$ with respect $p\in\mathcal{M}$ is called the \textit{cotangent bundle}. Thus, a tangent (co-tangent) bundle is a vectorial over structure upon the manifold; more in general
the disjoint union of the tensor in each point $p\in \mathcal{M}$ of a riemanian manifold
constitute the so called \textit{tensor bundle}. In Fig. \ref{tangentbundle} the columns represent the vector spaces built upon the flat manifold $\R ^2$: each column correspond to one tangent space, called \textit{fiber} of the tangent bundle while the flat manifold is the \textit{base} of the bundle
More in general the fiber can be constituted by an arbitrary vector space $V$ (as a space of tangent vector, covector, tensor spaces obtained as tensor product of them) and the \textit{vector bundle} associated has the \textit{local} structure of the Cartesian product of the manifold and the vector space: when this properties holds also globally the bundle is \textit{trivial}. An application with associate at each point $p$ of the base $\mathcal M$ a vector in the corresponding fiber $V_p\mathcal{M}$ is a \textit{section} of the fiber bundle. A \textit{vector field} over a manifold $\mathcal{M}$ can be regarded as a section
in some vector bundle $V\mathcal{M}$, as for instance the tangent bundle $T\mathcal{M}$.\\
In this case a vector field $\mathbf{X}\in\T\mathcal{M}$ that takes the value $\mathbf{X}(p)=X^i(p)\boldsymbol{\partial}_i$ in $p\in\mathcal{M}$ can be locally regarded as the tangent vector field of a curve $\gamma(t)\in\mathcal{M}$ with $\gamma(0)=p$ and $\dot{\gamma}(0)=\mathbf{X}(p)$.\\
The tangent bundle structure allows to introduce the operation of \textbf{tensor derivation}\footnote{The function over a manifold can be regarded as a tensor bundle of $(0,0)$-rank} $\mathcal{D}\mathcal{T}^{(r)}_{{s}}\rightarrow \mathcal{T}^{(r)}_{{s}}$ that in general have to satisfy the the following properties:
\begin{itemize}
\item Commutes with contractions, i.e. if $A\in\mathcal{T}^{(r)}_{{s}}$, $\{V_j\}_{j=1,\ldots,s}\in T\mathcal{}$ and $\{\alpha_i\}_{i=1,\ldots,s}$ 
\begin{equation}
\begin{split}
&\mathcal{D}A(\theta_{1},\ldots,\theta_{i},\ldots,\theta_{r},V_{1},\ldots,V_{j},\ldots,V_{s})=(\mathcal{D}A)(\theta_{1},\ldots,\theta_{i},\ldots,\theta_{r},V_{1},\ldots,V_{j},\ldots,V_{s})+\\
&+\sum_{i=0}^{r}A(\theta_{1},\ldots,\mathcal{D}\theta_{i},\ldots,\theta_{r},V_{1},\ldots,V_{j},\ldots,V_{s})+\\
&+\sum_{j=1}^{s}A(\theta_{1},\ldots,\theta_{i},\ldots,\theta_{r},V_{1},\ldots,\mathcal{D}V_{j},\ldots,V_{s})
\end{split}
\end{equation}
\item satisfy a Leibniz law with respect to tensor product, i.e.
$\mathcal{D}(A\otimes B)=\mathcal{D}A\otimes B+A \otimes \mathcal{D}B$.
\end{itemize}
From this two properties is follows that the action of a derivation is specified
by its action over function and vector field. The \textbf{Lie derivative} $\mathcal{L}_{\boldsymbol{X}}$ with respect to a vector field $\mathbf{X}$ generalize the concept of derivation along a curve: it measure the local rate of change of a tensor field along a curve whose tangent vector is $\mathbf{X}$.
Its action over functions $f$ and vector fields $Y\in T\mathcal{M}$ is given by:
\begin{equation}
\mathcal{L}_{\mathbf{X}}f:=\mathbf{X}(f)=X^{i}\partial_i f \qquad \mathcal{L}_{\mathbf{X}}\mathbf{Y}=[\mathbf{X},\mathbf{Y}]=\left(X^{j}\partial_{j}Y^{i}-Y^{j}\partial_{j}X^{i}\right)\boldsymbol{\partial}_i
\end{equation}
and consequently its action on one form is given by $\mathcal{L}_{\mathbf{X}}(\theta)=\left(\mathcal{L}_{\mathbf{X}}\theta\right)_i \D x^{i}=X^{i}\partial_{i}\theta_{j}+\theta_{j}\partial_{i}X^{i}$.\\
We notice that to calculate the values of the derivative of a tensor with respect to a vector field $\mathbf{X}$ in a point $p$ it is not sufficient to know its point-wise
value but it is required its local value for the presence of its derivatives\footnote{The only case where the point-wise information of $\mathbf{X}$ is sufficient is when it is applied to functions}. Moreover this derivation cannot be applied to a general vector bundle constructed over $\mathcal{M}$.  For this latest case is required to introduce the concept of \textbf{connection} over a vector bundle.
Let be $\Gamma(E)$ the set of all sections over a $q$-dimensional vector bundle $E$ with base $\mathcal{M}$ and fibre $V$ ; then a connection $\nabla$ over this vector bundle is a map $\nabla:\Gamma(E)\rightarrow \Gamma(T*\mathcal{M}\otimes E)$ such that
\begin{itemize}
\item the connection of a sum of sections is the sum of the connections of each section, i.e. $s_1,s_2\in \Gamma(E)$ it follows that $\nabla(s_1+s_2)=\nabla s_1+\nabla s_2$;
\item if $f$ is a function over $\mathcal{M}$ and $s\in\Gamma(E)$  then $\nabla (fs)=\D f\otimes s+f\nabla s$ where $\D f=\partial_{i}f \D x^{i}$ is the differential of the
function.
\end{itemize}
When the connection $\nabla s$ is contracted with a vector field $\mathbf{X}\in T\mathcal{M}$ is called \textbf{covariant derivative} of $s$ with respect to $\mathbf{X}$ and it is indicated with $\nabla_{\mathbf{X}}s$. The action of the connection is characterized by matrix valued $1$-form, called \textit{connection form} $\theta$: if $\mathbf{s}=s^{\alpha}\mathbf{e}_{\alpha}$ is a section in vector bundle $V$ whose local basis is $\{\mathbf{e}_{\alpha}\}_{\alpha=1,...,q}$
\begin{equation}
\nabla s=\D s^{\alpha} \otimes \mathbf{e}_{\alpha}+s^{\beta}\theta_{\beta}^{\alpha}\mathbf{e}_{\alpha}=\left(\partial_{i}s^{\alpha}+s^{\beta}\theta_{\beta;i}^{\alpha}\right)\D x^{i}\otimes \mathbf{e}_{\alpha}\,.
\end{equation}
Covariant derivative with respect to $\mathbf{X}=X^{i}\boldsymbol{\partial}_i$ on a tensor bundle $\mathcal{T}^{(r)}_{(s)}\mathcal{M}$ acts like a derivation; it follows that also in this case it is uniquely defined by its action on function and vectors
\begin{equation}
\nabla_{\mathbf{X}}f=X^{i}\partial_{i}f \qquad \nabla_{\boldsymbol{X}}\mathbf{Y}=X^{j}\left(\partial_{j}Y^{i}+Y^{k}\theta^{i}_{k;j}\right)\boldsymbol{\partial}_i
\end{equation}
and consequently, in the case of the covariant derivative with respect to $\mathbf{X}$
of a general tensor $t=t^{i_1 \ldots i_s}_{j_1 \ldots j_r} \boldsymbol{\partial}_{i_1}\otimes \ldots \boldsymbol{\partial}_{i_s}\otimes \D x^{j_1} \otimes \ldots \D x^{j_r}$
it is given by
\begin{equation}
(\nabla_{\mathbf{X}} t)^{i_1 \ldots i_s}_{j_1 \ldots j_r}=X^{k}\left(\partial_k t^{i_1 \ldots i_s}_{j_1 \ldots j_r} + \sum_{l=1}^{r}\theta^{i_l}_{i_m;k} t^{i_1 \ldots i_m \ldots i_s}_{j_1 \ldots j_r}-\sum_{l=1}^{s}\theta^{j_m}_{j_l;k} t^{i_1 \ldots i_m \ldots i_s}_{j_1 \ldots j_r} \right)
\end{equation}
For connections on the tangent bundle, the \textbf{torsion} $(2,1)$-rank tensor is
defined as
\begin{equation}
\mathrm{Tor}(\mathbf{X},\mathbf{Y})=\nabla_{\mathbf{X}}\mathbf{Y}-\nabla_{\mathbf{Y}}\mathbf{X}-[\mathbf{X},\mathbf{Y}]
\end{equation}
and in components reads
\begin{equation}
\mathrm{Tor}_{ij}^{k}=\theta_{i;j}^{k}-\theta_{j;i}^{k}-\left([\boldsymbol{\partial}_i,\boldsymbol{\partial}_j]\right)^k\,.
\end{equation}
In general $\nabla_{\mathbf{X}}\nabla_{\mathbf{Y}}\neq\nabla_{\mathbf{Y}}\nabla_{\mathbf{X}}$ a measure of the difference between these derivatives is the \textbf{curvature} $R$ of the connection, i.e.
\begin{equation}
R(\mathbf{X},\mathbf{Y})=\nabla_{\mathbf{X}}\nabla_{\mathbf{Y}}-\nabla_{\mathbf{Y}}\nabla_{\mathbf{X}}-\nabla_{[\mathbf{X},\mathbf{Y}]}
\end{equation}

\section{Differential forms, exterior differentiations, integration of forms}

Let us indicate with $A^{k}(\mathcal{M})$ the tensor bundle constructed over the manifold $\mathcal{M}$ as a disjoint union of the $k$-form spaces $\Lambda^{k}_{p}$. For the $n$-dimensional manifold $\mathcal{M}$ the space $A(\mathcal{M})=\sum_{k=0}^{n}A^{k}(\mathcal{M})$ is a graded algebra as with respect to the operation of sum $+$ and wedge product $\wedge$ of differential forms, meaning that $A(\mathcal{M})$ is a direct sum of a sequence of a vector spaces, and the product $\wedge$ defines a map $\wedge:A^{r}(\mathcal{M})+A^{s}(\mathcal{M})\rightarrow A^{r+s}(\mathcal{M})$ where $A^{r+s}(\mathcal{M})$ is zero when $r+s>n$.
The \textbf{exterior derivative} $\D$ is a derivation introduced on $A(\mathcal{M})$
such that $\D(A^{k}(\mathcal{M}))\subset A^{k+1}(\mathcal{M})$ such that for any $\alpha,\beta \in A(\mathcal{M})$:
\begin{itemize}
\item $\D (\alpha+\beta)=\D \alpha +\D \beta$;
\item if $\alpha\in A^{r}(\mathcal{M})$ then $\D(\alpha\wedge \beta)=\D \alpha \wedge \beta+(-1)^{r} \alpha \wedge \D \beta$;
\item if $f$ is a smooth function $\mathcal{M}$ (i.e. $f\in A^{0}(\mathcal{M})$), where $\D f= \partial_i f \D x^{i}$
\item  if $f$ is a smooth function $\mathcal{M}$ then $\D(\D f)=0$.
\end{itemize}
The exterior differentiation of a differential $k$-form $\alpha \in A^{k}(\mathcal{M})$
is given by
\begin{equation}
\begin{split}
&\left(\D \alpha\right) \left(\mathbf{X}_1,\ldots,\mathbf{X}_{k+1}\right)=\sum_{i=1}^{k+1}(-1)^{i+1}\mathbf{X}_{i}\left(\alpha(\mathbf{X}_1,\ldots,\mathbf{X}_{i-1},\mathbf{X}_{i+1},\ldots\mathbf{X}_{k+1})\right)+\\
&+\sum_{1\leq i<j\leq k+1} (-1)^{i+j}\alpha([\mathbf{X}_i,\mathbf{X}_j],\mathbf{X}_1, \ldots \mathbf{X}_{i-1} \mathbf{X}_{i+1} \ldots \mathbf{X}_{j-1}\mathbf{X}_{j+1}\ldots\mathbf{X}_{k+1})
\end{split}
\end{equation}
The \textit{Poincare's Lemma} states that $\D^2 \alpha=0$ for all $\alpha\in A(\mathcal{M})$. The space $A(\mathcal{M})$ endowed with the differential $\D$ is called a complex.
In any subspace $A^{k}(\mathcal{M})$ two subspaces can be identified:
\begin{itemize}
\item the space $Z^{k}(\mathcal{M})$ of all $k$-forms $\alpha$ that are \textit{closed}, so
$\D \alpha=0$, (also called \textit{cocycles})
\item the space $B^{k}(\mathcal{M})$ of all $k$-forms $\alpha$ that are \textit{exact}, so
that exist $\beta\in C^{k-1}(\mathcal{M})$ such that $\D \beta=\alpha$, (also called \textit{coboundaries}).
\end{itemize}
The cohomology space $H^{k}(\mathcal{M};\mathbb{R})$ is defined as
\begin{equation}
H^{k}(\mathcal{M};\mathbb{R})=Z^{k}(\mathcal{M})/B^{k}(\mathcal{M})
\end{equation}
and an element $[\omega]$ in this space is the \textit{equivalence class} of $k$-forms that differ among them for an exact form, i.e. $[\omega_1]=[\omega_2]$ if $\omega_1-\omega_2=\D phi$. From the definition of differential $k$-form it follows that $H^{k}(\mathcal{M},\mathbb{R})=0$ if $k > n$.\\
The \textbf{De Rham cohomology space} $H^{*}(\mathcal{M};\mathbb{R})$ is a graded space obtained as the direct sum of the $k$-dimensional cohomology spaces:
\begin{equation}
H^{*}(\mathcal{M},\mathbb{R})=\otimes_{k\geq 0} H^{k}(\mathcal{M};\mathbb{R}) \,.
\end{equation}
and it is a multiplication ring with the the addition $[\omega_1]+[\omega_2]=[\omega_1 +\omega_2]$ and the multiplication $[\omega_1]\smile [\omega_2]=[\omega_1 \wedge \omega_2]$.
Moreover if $f:\mathcal{M}^{n}\rightarrow \mathcal{N}^{k}$ is a smooth map the pullback\footnote{Let be $h:\mathcal{N}\rightarrow \mathcal{M}$ an application between two manifolds, and let $f:\mathcal{M}\rightarrow \mathbb{R}$ be a function over $\mathcal{M}$, then the \textbf{pullback} of a function is defined as the function $(h^{*}f):\mathcal{N}\rightarrow\mathbb{R}$ s.t. $(h^{*}f)=f\circ h$. The wedge product and exterior differentiation commutates with the pullback of forms, i.e. $h^{*}(\omega_1 \wedge \omega_2)=h^{*}\omega_1 \wedge h*\omega_2$ and $\D (h^{*}f)=h^{*}\D\omega$.} $f^{*}:A^{p}(\mathcal{N}^{k})\rightarrow A^{k}(\mathcal{M}^{n})$ is a \textit{homomorphism induced by the mapping} among cohomology rings, i.e. $f^{*}:H^{*}(\mathcal{N}^{k};\mathbb{R})\rightarrow H^{*}(\mathcal{M}^{k};\mathbb{R})$ with $f^{*}[\omega]=[f^{*}\omega]$.\\
Two smooth mapping $f_{0}:\mathcal{M}^n\rightarrow\mathcal{N}^k$ and $f_{1}:\mathcal{M}^n\rightarrow\mathcal{N}^k$ are smoothly homotopic if there is a smooth \textit{homotopy} mapping $F:\mathcal{M}^n\times[0,1]\rightarrow \mathcal{N}^k$ such that $F(x,k)=f_{k}(x)$ with $k=0,1$. In this case the action of the homomorphism of cohomology rings $f^{*}_{0}:H^{*}(\mathcal{N}^{k};\mathbb{R})\rightarrow H^{*}(\mathcal{M}^{n};\mathbb{R})$ and $f^{*}_{1}:H^{*}(\mathcal{N}^{k};\mathbb{R})\rightarrow H^{*}(\mathcal{M}^{n};\mathbb{R})$
coincides. Two manifolds $\mathcal{M}^{n}$ and $\mathcal{N}^{n}$ are said to be \textit{homotopically equivalent} if there exists two mappings $f:\mathcal{M}^{n}\rightarrow\mathcal{N}^{k}$ and $g:\mathcal{N}^{k}\rightarrow\mathcal{M}^{n}$ such that $fg$ and $gf$ are homotopic to the identity. It can be proved that two homotopically equivalent manifolds have isomorphic cohomology groups.\\
The integration of exterior derivatives allows to connect local and global properties of a manifold. The first step consists in defining an \textbf{orientable manifold} as $n$-dimensional manifold $\mathcal{M}$ where there exists a continuous and nonvanishing exterior differential $n$-form; two differential $n$-forms which differ everywhere by a function factor which is always positive define an \textit{orientation}. Let us suppose that $\mathcal{M}$ is a manifold oriented by a differential $n$-form $\omega$, then a chart $(U,x)$ is \textit{compatible} if $\omega$ and $\D x^{1}\wedge\ldots\wedge \D x^{n}$ define the same orientation.\\
The main results that allows to define the integration over a manifold is the \textbf{Partition of Unity Theorem}: suppose $\Sigma$ is an open covering of a smooth manifold $\mathcal{M}$. Then there exists a family of smooth functions $\{g_{\alpha}\}$ on $\mathcal{M}$ satisfying the following conditions:
\begin{itemize}
\item $0\leq g_{\alpha} \leq 1$ and the $\mathrm{supp} g_{\alpha}=\{p\in\mathcal{M}|g_{\alpha}(p)\neq 0\}$ is compact for each $\alpha$. Moreover, there exists an open set $W_i\in\Sigma$ such that $g_{\alpha}\subset W_{i}$;
\item For each point $p\in\mathcal{M}$ there is a neighbourhood $U$ that intersects $\mathrm{supp} g_{\alpha}$ for only a finite number of $\alpha$
\item $\sum_{\alpha} g_{\alpha}=1$.
\end{itemize}
The \textbf{integral} of a differential $m$-form $\omega$ is defined as 
\begin{equation}
\int_{\mathcal{M}} \omega=\sum_{\alpha}\int_{W_i} g_{alpha} \omega
\end{equation}
where in a local coordinate system $g_{alpha} \omega=f(x^1,\ldots,x^{n})\D x^{1}\wedge\ldots\wedge\D x^{n}$ so that $\int_{W_i} \,\, g_{alpha} \omega = \int_{W_i}f(x^1,\ldots,x^{n})\D x^{1}\ldots\D x^{n}$. Let us suppose that $h:\mathcal{N}\rightarrow \mathcal{M}$ is the imbedding of a $k$-dimensional submanifold in the $n$-dimensional ambient manifold $\mathcal{M}$. Then the integral of the differential $k$-form $h^{*}\omega$ over $\mathcal{N}$ is defined as
\begin{equation}
\int_{h(\mathcal{N})} \omega=\int_{\mathcal{N}}h^{*}\omega\,\,\,.
\end{equation}
A very relevant result in integration theory of differential forms is the \textbf{Stokes Theorem}: let $\omega$ be a $(n-1)$-form and $D\subseteq \mathcal{M}$ and with a smooth or piecewise smooth boundary $\partial D$ then
\begin{equation}
\int_{\partial D}\omega=\int_{D}\D\omega\,\,\,.
\end{equation}
This formula has its importance as it allows to characterize the topology of a certain domain $D$ over a manifold $\mathcal{M}$ as it establishes a duality (Poincaré duality) among the boundary operator $\partial$ and the coboundary operator (i.e. the exterior differentiation) $\D$ on forms. This result can be expressed
\begin{equation}
\label{eq:Poincare_Duality}
\int_{\partial D}\omega=(\partial D,\omega)=(D, \D \omega)=\int_{D}\D\omega
\end{equation}
which represents the link between homology, the rigorous mathematical way to classify the manifolds with respect to their "holes", and the De Rham cohomology of differential forms on manifolds.\\
A geometric \textit{cycle} on a manifold $\mathcal{M}^n$ is a pair $(\mathcal{N}^k,f)$ consisting of a smooth mapping $f:\mathcal{N}^k \rightarrow \mathcal{M}^n$. If there is an orientation on the manifold $\mathcal{N}^k$, then such a cycle will be said to be oriented.
Any closed $k$-form $\omega$ on a manifold $\mathcal{M}^n$ specifies a function on the set of all $k$-dimensional cycles $z=(\mathcal{N}^k,f)$ by the formula
\begin{equation}
\left\langle \omega ,z \right\rangle=\int_{N_k} f^{*}\omega \,.\,.
\end{equation} 
It can be proved that the value of $\left\langle \omega ,z \right\rangle$ depends only by
the cohomology group $[\omega]$.\\
Let $A_{p}(\mathcal{M})$ the space generated by all the oriented $p$-cycle over the field $\mathbb{R}$. For each element $z$ of this space we can define the integral of a closed $p$-form $\omega$ over $z$. Namely let $z=\lambda_1 z_1 +\ldots+\lambda_k z_k$ where $\lambda_i\in\mathbb{R}$ and $z_{i}=(\mathcal{N}_{i}^p,f_i)$ are ordinary cycles. Then 
\begin{equation}
\left\langle \omega, z \right\rangle=\sum_{i}\lambda_{i}\left\langle \omega,z_i\right\rangle=\sum_{i}\lambda_i \int_{\mathcal{N}_i} f_{i}^{*}\omega
\end{equation}
Let $B_{p}(\mathcal{M})\subset A_{p}(\mathcal{M})$ be the subspace that consist of the cycles such that the integral of all closed $p$-forms over these cycles vanish. The quotient 
space 
\begin{equation}
H_{p}(\mathcal{M}^p;\mathbb{R})=A_{p}(\mathcal{M}^n)/B_{p}(\mathcal{M}^n)
\end{equation}
is called the real $p$-dimensional homology group of the manifold $\mathcal{M}^n$.
The homology space is dual to the cohomology group $H^{p}(\mathcal{M}^p;\mathbb{R})$
and for any nonzero element $[\omega]\in H^{p}(\mathcal{M}^p;\mathbb{R})$ the linear functional
\begin{equation}
\left\langle [\omega],[z]\right\rangle=\left\langle \omega,z\right\rangle \qquad z\in H_{p}(\mathcal{M}^p;\mathbb{R})\,.
\end{equation}
For compact manifolds all the cohomology groups are finite-dimensional and their dimensions are called \textit{Betti numbers} $b_p(\mathcal{M})=\mathrm{dim}H_{p}(\mathcal{M}^p;\mathbb{R})=\mathrm{dim} H^{p}(\mathcal{M}^p;\mathbb{R})$ and are topological invariants.
The alternating sum of the Betti numbers is the \textit{Euler characteristics}
$\chi(\mathcal{M})=\sum_{i=0}^{n} (-1)^{i}\,b_{i}(\mathcal{M})$; as for a compact $n$-dimensional connected manifold without boundaries the Poincare duality implies $H^{n-k}(\mathcal{M}^p;\mathbb{R})=H^{k}(\mathcal{M}^p;\mathbb{R})$ it follows that for odd-dimensional manifolds the Euler characteristic vanishes $\chi(\mathcal{M}^{2n+1})=0$.
\section{Riemannian structure}
Thanks to the vector structure, a scalar product, can be defined on $T_p\mathcal{M}$ as $g_p:T_p\mathcal{M} \times T_p\mathcal{M}\rightarrow\R$. This is a way to associate a \textit{real} number to any pair $(\mathbf{X}_p,\mathbf{Y}_p)$ of tangent vectors. This product inherits all the features of the canonical scalar product on the linear space $\R ^n$; in particular $\|\mathbf{X}_p\|_{g_p}^2=g_p(\mathbf{X}_p,\mathbf{X}_p)>0$ for all tangent vector $\mathbf{X}_p$ and $g_p(\mathbf{X}_p,\mathbf{Y}_p)=0$, if and only if one of the tangent vectors $X_p$ or $Y_p$ is null.\\
The Riemannian structure allows to define the \textit{gradient} $\mathbf{grad}_{g}f$ of a function $f$ on $\mathcal{M}$ as the vector field with the property
\begin{equation}
g(\mathbf{grad}_g f, \mathbf{Y})=\mathbf{Y}(f)=Y^{i}\partial_i f\,.
\end{equation}
and the \textit{divergence} of a vector field $\mathbf{X}$ as
\begin{equation}
\left(\mathrm{div}_{g}\mathbf{X}\right)=\sum_{j=1}^N \,\,\D x^{j}\left(\nabla_{\boldsymbol{\partial}_j}\mathbf{X}\right)\,\,.
\end{equation}
If this properties holds for every point $p\in\mathcal{M}$ for the tensor field $(0,2)$-rank $g=g_{ij}\D x^{i}\otimes \D x^{j}$ such a tensor is a \textbf{metric tensor field} that defines a \textbf{Riemannian structure}
over the manifold $\mathcal{M}$.\\
The definition of a metric tensor allows to define the length of a curve $c(t):[a,b]\rightarrow\mathbb{R}$ as 
\begin{equation}
L(c)=\int_{a}^{b}\|\dot{\mathbf{c}}(t)\|^2 \D t \,.
\end{equation}
and the volume form $\D \mathrm{Vol}_g$
\begin{equation}
\D \mathrm{Vol}_g=|\mathrm{det}g_{ij}|^{1/2}\D x^{1}\wedge\ldots\wedge \D x^{N}\,.
\end{equation}
The inverse of $g_{ij}$ is given by the $g^{kl}$ matrix with $g_{ij}g^{jk}=\delta_{i}^{k}$; this two matrices allows to define the raising and lowering of indices of a tensor, i.e.
\begin{equation}
g_{k i_{l}}t_{\quad\quad\quad j_1 \ldots j_r}^{i_1\ldots i_l \ldots i_r}=t_{\quad\quad\quad k \quad\quad\quad j_1 \ldots j_r}^{i_1\ldots i_{l-1}\quad i_{l+1} \ldots i_r} \qquad\qquad g^{k j_{l}}t_{\quad\quad j_1 \ldots j_l \ldots j_r}^{i_1\ldots i_r}=t_{\quad\quad j_1\ldots j_{l-1}\quad j_{l+1}\ldots j_r}^{i_1\ldots i_r\quad\quad\quad k}
\end{equation}
This allows to define the \textit{metric-dependent trace of a  $(0,2)$-rank tensor} $A=A_{ij}\D x^{i}\otimes\D x^{j}$ as:
\begin{equation}
\mathrm{Tr}^g\left(A\right)=g^{ij}A_{ij}=A^{i}_{\quad i}\,.
\end{equation}
The Riemannian structure puts some constraints on the definition of a connection on the tangent and tensor bundles over a given manifold; in particular, a \textbf{Levi-Civita connection} is \textit{compatible with the metric}, that means
\begin{equation}
\left(\nabla_{\mathbf{X}} g \right)_{ij}= X^{k} \partial_{k}g_{ij}-\theta^{l}_{i;k}g_{lj}-\theta^{l}_{j;k}g_{il}
\end{equation}
and \textit{torsion free}, i.e. $\mathrm{Tor}=0$. In this case the components of the matrix-valued connection form are expressed by the \textit{Christoffel symbol}
$\Christoffel{k}{i}{j}=\theta^{k}_{j;i}$. It can be proved that the for a given metric $g$ over $\mathcal{M}$ a unique Levi-Civita is defined.\\
The Christoffel symbol can be expressed as a function of derivatives of the components of the metric tensor
\begin{equation}
\Christoffel{i}{j}{k}=\dfrac{1}{2}g^{il}\left(\partial_{j}g_{lk}+\partial_{k}g_{jk}-\partial_{l}g_{jk}\right)\,.
\end{equation}
The  curvature tensor for a Levi-Civita connection specified by the symbols $\Christoffel{i}{j}{k}$ is the \textbf{Riemann curvature tensor} that in components reads:
\begin{equation}
R^{i}_{jkl}=\partial_{l}\Christoffel{i}{j}{k}-\partial_{j}\Christoffel{i}{k}{l}+\Christoffel{m}{j}{k}\Christoffel{i}{m}{l}-\Christoffel{m}{j}{l}\Christoffel{i}{m}{k}
\end{equation}
and whose completely covariant version is the tensor $R_{ijkl}=g_{mi}R^{i}_{jkl}$
with basic symmetries:
\begin{equation}
R_{ijkl}=-R_{jikl}=-R _{ijlk}=R_{klij}\,.
\end{equation}
In order to give a geometrical interpretation of Riemann curvature, let us introduce the concept \textit{geodesic curve} as the shortest smooth curve that connects two points on a Riemannian manifold; a variational formulation of this condition for a geodesic $\gamma=(x^{1}(t)\ldots x^{N}(t)))$ is given by:
\begin{equation}
\Delta_{\dot{\boldsymbol{\gamma}}}\dot{\boldsymbol{\gamma}}=0 \quad\Rightarrow\quad \ddot{x}^{i} +\Christoffel{i}{k}{j} \ddot{x}^{j}\ddot{x}^{k}\,.
\end{equation}
Now let us consider a smooth one parameter family of geodesic $\gamma_{\tau}$ with $\gamma_{0}(t)=\gamma(t)$ than the \textit{Jacobi vector field} $\mathbf{J}$ along the geodesic $\gamma(t)$ is defined as
\begin{equation}
\mathbf{J}=\partial \gamma_{\tau}(t)\Bigr|_{\tau=0}
\end{equation}
The Riemann curvature tensor gives a measure of the local geodesic spread. The evolution of the Jacobi field along a given geodesic is described by the \textit{Jacobi-Levi-Civita equation}. Let us suppose that along the geodesic it is consider a orthonormal frame $\{\mathbf{e}_{i}\}_{i=1,...,N}$ in $T_{\gamma(t)} \mathcal{M}$, parallel transported all along $\Delta \mathbf{e}_{i}=0$  and with $\mathbf{e}_{1}=\dot{\boldsymbol{\gamma}}/\|\dot{\boldsymbol{\gamma}}\|_g$ than if the jacobi vector field is expressed $\boldsymbol{J}=y^{i}\mathbf{e}_{i}$ the Jacobi Levi Civita for the geodesic spread reads
\begin{equation}
\ddot{y}^k+\|\dot{\boldsymbol{\gamma}}\|^2_{g} \mathrm{R}^{k}_{1j1} y^{j}=0 \,.
\end{equation}
The contraction $C_{1,2}(R)$ of the Riemann curvature tensor is the \textbf{Ricci curvature tensor}
\begin{equation}
\mathrm{Ric}_{ij}=R^{k}_{ikj}=\partial_{l}\Christoffel{l}{j}{i}-\partial_{j}\Christoffel{l}{l}{i}+\Christoffel{l}{l}{k}\Christoffel{k}{j}{i}-\Christoffel{l}{j}{k}\Christoffel{k}{i}{l}\,.
\end{equation}
For a given metric $g$ the Ricci curvature gives a local measure of the difference between the volume form with respect to an euclidean metric $g_{\mathbb{E}}$, i.e. if in a certain point $g_{ij}=\delta_{ij}+o\left(\|\mathbf{X}\|_{g}^2\right)$ with $\mathbf{X}=X^{i}\boldsymbol{\partial}_{i}$
\begin{equation}
\D \mathrm{Vol}_g=\D \mathrm{Vol}_{g_{\mathbb{E}}}\left(1-\dfrac{1}{6}\mathrm{Ric}_{jk}X^{k}X^{j}+o\left(\|\mathbf{X}\|_{g}^3\right)\right)
\end{equation}
so in the directions $\mathbf{X}$ such that $\mathrm{Ric}(\boldsymbol{X},\boldsymbol{X})$ is positive the volume is contracted with respect to the Euclidean volume.\\
The metric-dependent trace of the Ricci curvature tensor is the \textbf{Scalar curvature} $\mathcal{R}=\mathrm{Tr}^g \mathrm{Ric}=g^{ij}\mathrm{Ric}_{ij}$.

\section{Riemmanian geometry of codimension one submanifolds (regular level sets)}

Let $(\mathcal{M},\bar{g})$ be a $N$-dimensional Riemannian manifold  whose Levi-Civita commection is $\overline{\nabla}$. A \textbf{regular Submanifold} of dimension $K$ is a subset $\Sigma$ such that for every point $p\in \Sigma$ such that for a certain chart $(U,x)$ over $\mathcal{M}$ it holds $x\left(U \cap Sigma\right)=x(U)\cup {\mathbb{R}^{K} \times \{c\}}$ for some $c\in\mathbb{R}^{N-K}$, that means that exists a chart over $\mathcal{M}$ such that for its restriction over $\Sigma$ there are $N-K$ fixed components. In particular, 
the set defined as a locus that in a chart $(U,x)$ reads $\Sigma=\left\{p\in\mathcal{M}| f_{A}(x^1(p),...,x_N(p))\right\}=0\qquad \text{for}\,\, A=1,...,L$ it is a $K$-dimensional level sets (with $K\leq L$) if the rank of the Jacobian $\partial_{i}f_{A}$ is $K$.\\
The equipotential level sets discussed in the first part of this manuscript are an example of regular submanifold without boundaries in absence of critical point of potential energy.
Moreover in this specific case, we have a local \textbf{regular foliation} $\mathcal{F}$ of the configuration space, namely when the ambient manifolds can be regarded as the disjoint union of connected regular submanifolds $\Sigma_{\mathbf{a}\in \mathbb{R}^{N-K}}$ called \textit{leaves} of dimension $K$ (and co-dimension $N-K$) such that in a neighbourhood of any point of the ambient space exists a chart $(U,x)$ such where the coordinate can be expressed in the form $(a^{1},...,a^{N-K},x^1,...,x^{K})$. In what follows we consider level sets of one single function so that $K=N-1$.\\
As we supposed that the ambient space has a Riemannian structure, each leaf inherits
a metric structure, the so called \textbf{First  Fundamental Form} $\mathrm{I}:T_{p}\Sigma\times T_{p}\Sigma\rightarrow\mathbb{R} $ defined over a regular submanifold of co-dimension one
\begin{equation}
\mathrm{I}_{\Sigma}(\mathbf{X},\mathbf{Y}):\equiv \bar{g}(\mathbf{X},\mathbf{Y})\qquad \mathbf{X},\mathbf{Y} \in T_{p}\Sigma
\end{equation}
and it coincides with the restriction of metric tensor $\bar{g}$ on the immersed co-dimension one submanifold $\Sigma$. This allows to define a Levi-Civita connection $\nabla$ over the regular submanifold. In order to introduce a more suitable notation for a fixed $\Sigma$, we will use in what follows $g=\mathrm{I}$.\\
Moreover, for a co-dimension one regular submanifold $\Sigma$ in a ambient space $\mathcal{M}$ it is possible to define a \textbf{normal vector field} as the vector
field $\boldsymbol{\nu}\neq \mathbf{0}$ such that $\bar{g}(\boldsymbol{\nu},\mathbf{X})=0$ for every  $\mathbf{X}\in T_{p}\Sigma$.\\
The rate of (covariant) variation of the normal vector field in a direction tangent to the submanifolds defines is intuitively related with the concept of curvature for a
immersed submanifold. This is can be formalized as follows. First of all let us consider the \textbf{Weingarten operator or Shape Operator} $W_{\boldsymbol{\nu}_{p}}:T_{p}\Sigma\rightarrow T_{p}\Sigma$:
\begin{equation}
W_{\boldsymbol{\nu}_{p}}(X):\equiv \overline{\nabla}_{\mathbf{X}}\boldsymbol{\nu}\bigr |_{p} \qquad \mathbf{X}\in T_{p}\Sigma\,\,\,\,\,.
\end{equation}
This operator can be regarded as $(1,1)$-rank tensor field over he submanifold tangent space; the induced metric structure over the submanifold allows to construct a $(0,2)$-rank tensor field over this submanifold called \textbf{Second Fundamental Form} $\mathrm{II}_{\Sigma}:T_{p}\Sigma\times T_{p}\Sigma\rightarrow\mathbb{R} $
\begin{equation}
\mathrm{II}_{\Sigma}(\mathbf{X},\mathbf{Y})=\bar{g}(W_{\nu}(\mathbf{X}),\mathbf{Y})= \bar{g}( \nabla_{\mathbf{X}}\boldsymbol{\nu},\mathbf{Y})=- \bar{g}(\boldsymbol{\nu}, \nabla_{\mathbf{X}}\mathbf{Y})\qquad \mathbf{X},\mathbf{Y}\in T_{p}\Sigma
\end{equation}
The second fundamental form is symmetric in its arguments as:
\begin{equation}
\begin{split}
\mathrm{II}(\mathbf{X},\mathbf{Y})&= \bar{g}( \overline{\nabla}_{\mathbf{X}}\boldsymbol{\nu},\mathbf{Y})=-\bar{g
}(\boldsymbol{\nu}, \overline{\nabla}_{\mathbf{X}}\mathbf{Y})=-\bar{g}(\boldsymbol{\nu}, 
\overline{\nabla}_{\mathbf{Y}}\mathbf{X})-\bar{g}(\boldsymbol{\nu},\left[\mathbf{X},\mathbf{Y}\right])=\\
&=\bar{g}( \overline{\nabla}_{\mathbf{Y}}\boldsymbol{\nu},\mathbf{X})=\mathrm{II}(\mathbf{Y},\mathbf{X})\qquad \mathbf{X},\mathbf{Y}\in T_{p}\Sigma
\end{split}
\end{equation}
The eigenvalues $(\lambda_1,...,\lambda_n)$ of the Weingarten operator are called \textbf{principal curvatures} and the metric-dependent trace of shape operator is called \textbf{mean curvature}:
\begin{equation}
\label{eq:def_Meancurvature}
h_{\Sigma,\bar{g}}=\dfrac{\sum_{i=1}^{N-1}\lambda_{i,\bar{g}}}{N-1}=\dfrac{\mathrm{Tr}^{g}(\mathrm{II})}{N-1}=\dfrac{\tau_{1,\bar{g}}}{N-1}
\end{equation}
We derive a formula for the variation of mean curvature along the normal direction, in a coordinate system $(x^{0},x^{1},...,x^{N-1})$, where $\boldsymbol{\partial}_0=\boldsymbol{\nu}$, for a co-dimension one regular foliation
\begin{equation}
\label{eq:AppGeo_LieNu_TraceIIform}
\mathcal{L}_{\boldsymbol{\nu}}(\tau_{1,\bar{g}})=\overline{\nabla}_{\boldsymbol{\nu}}\left(g^{ij}\mathrm{II}_{ij}\right)=\mathcal{L}_{\boldsymbol{\nu}}(g^{ij})\mathrm{II}_{ij}+g^{ij}\mathcal{L}_{\boldsymbol{\nu}} \left(\mathrm{II}_{ij}\right) \qquad i,j=1,...,N-1
\end{equation}
We note that
\begin{equation}
\label{eq:AppGeo_LieNu_InvMetric}
\mathcal{L}_{\boldsymbol{\nu}}\left(g^{ij}g_{jk}\right)=0
\end{equation}
from what follows
\begin{equation}
\mathcal{L}_{\boldsymbol{\nu}}g^{ij}=-g^{jk}g^{il}\mathcal{L}_{\boldsymbol{\nu}}(g_{kl})\,.
\end{equation}
The last term can be calculated considering that the action of Lie derivative and covariant differentiation coincide on functions
\begin{equation}
\label{eq:AppGeo_LieNu_InvMetric_2}
\mathcal{L}_{\boldsymbol{\nu}}(g_{kl})=\overline{\nabla}_{\boldsymbol{\nu}}\left(\bar{g}(\mathbf{e}_{k},\mathbf{e}_{l})\right)=2(\overline{\nabla}_{\boldsymbol{\nu}}^{\bar{g}}\mathbf{e}_{k},\mathbf{e}_{l})=2(\overline{\nabla}_{\mathbf{e}_{k}}\boldsymbol{\nu},\mathbf{e}_{l})=2\mathrm{II}(\mathbf{e}_{k},\mathbf{e}_{l})=2\mathrm{II}_{kl}
\end{equation}
so that substituting the last expression in \eqref{eq:AppGeo_LieNu_InvMetric} we obtain
\begin{equation}
\label{eq:AppGeo_LieNu_InvMetric_3}
\mathcal{L}_{\boldsymbol{\nu}}g^{ij}=-2\mathrm{II}^{ij}
\end{equation}
and the first term of right-side in eq. \eqref{eq:AppGeo_LieNu_TraceIIform} results
\begin{equation}
\mathcal{L}_{\boldsymbol{\nu}}(g^{ij})\mathrm{II}_{ij}=-2\mathrm{II}^{ij}\mathrm{II}_{ij}=-2(\mathrm{II}^2)^{i}_{i}=-2\mathrm{Tr}^{g}\left(\mathrm{II}^2\right)\left(=-2 \sum_{i=1}^{N-1}\lambda_{1,\bar{g}}^2:=-2 \tau_{2,\bar{g}}\right)
\end{equation}
Let us now consider the Lie derivative of second fundamental form along the normal field.
\begin{equation}
\label{eq:geoApp_LieNu_2ndForm}
\begin{split}
\mathcal{L}_{\boldsymbol{\nu}}\mathrm{II}_{ij}&=\mathcal{L}_{\boldsymbol{\nu}}\mathrm{II}(\mathbf{e}_i,\mathbf{e}_j)=\overline{\nabla}_{\boldsymbol{\nu}}\bar{g}(\nabla_{\mathbf{e}_i}^{\bar{g}}\boldsymbol{\nu},\mathbf{e}_j)=\bar{g}(\overline{\nabla}_{\boldsymbol{\nu}}^{\bar{g}}\overline{\nabla}_{\mathbf{e}_i}\boldsymbol{\nu},\mathbf{e}_j)+\bar{g}(\overline{\nabla}_{\mathbf{e}_i}\boldsymbol{\nu},\overline{\nabla}_{\mathbf{\nu}}\boldsymbol{e}_j)=\\
&=\bar{g}\left(\bar{R}(\boldsymbol{\nu},\mathbf{e}_i)\boldsymbol{\nu},\mathbf{e}_j\right)+\bar{g}(\overline{\nabla}_{\mathbf{e}_i}\overline{\nabla}_{\boldsymbol{\nu}}\boldsymbol{\nu},\mathbf{e}_j)+\bar{g}(\left[\boldsymbol{\nu},\mathbf{e}_i\right],\mathbf{e}_j)+
\bar{g}(\overline{\nabla}_{\mathbf{e}_i}\boldsymbol{\nu},\overline{\nabla}_{\boldsymbol{\nu}}\mathbf{e}_j)\\
\end{split}
\end{equation}
where $R$ is the Riemannian curvature tensor of the ambient space. Under the hypothesis that $\overline{\nabla}_{\boldsymbol{\nu}}\boldsymbol{\nu}=0$ and using the antisymmetrical properties of the Riemann tensor we obtain:
\begin{equation}
\label{eq:geoApp_LieNu_2ndForm_2}
\begin{split}
&\mathcal{L}_{\boldsymbol{\nu}}\mathrm{II}_{ij}=-\bar{g}\left(\bar{R}(\mathbf{e}_i,\boldsymbol{\nu})\boldsymbol{\nu},\mathbf{e}_j\right)+
\bar{g}(\overline{\nabla}_{\mathbf{e}_i}\boldsymbol{\nu},\overline{\nabla}_{\mathbf{e}_j}\boldsymbol{\nu})=-\bar{R}_{j0i0}+\mathrm{W}^{k}_{i}\mathrm{W}^{l}_{j}\bar{g}(\mathbf{e}_{k},\mathbf{e}_{l})=\\
&=-\bar{R}_{j0i0}+\mathrm{II}^{k}_{i}\mathrm{II}^{l}_{j}\bar{g}_{k,l}=-
\bar{R}_{j0i0}+\mathrm{II}_{i}^{k}\mathrm{II}_{kj}\,.
\end{split}
\end{equation}
Putting together eqs. \eqref{eq:AppGeo_LieNu_InvMetric} and \eqref{eq:geoApp_LieNu_2ndForm_2} in eq.\eqref{eq:AppGeo_LieNu_TraceIIform} we obtain
\begin{equation}
\begin{split}
&\mathcal{L}_{\boldsymbol{\nu}}(\tau_{1,\bar{g}})=-2\mathrm{II}^{ij}\mathrm{II}_{ij}+g^{ij}\left(-\overline{R}_{j0i0}+\mathrm{II}_{i}^{k}\mathrm{II}_{kj}\right)=
-\left(\mathrm{II}^{ij}\mathrm{II}_{ij}+\overline{\mathrm{Ric}}_{00}\right)=\\
&=-\mathrm{Tr}^{g}(\mathrm{II}^2)-\overline{\mathrm{Ric}}(\boldsymbol{\nu},\boldsymbol{\nu})=-\tau_{2,\bar{g}}-\overline{\mathrm{Ric}}(\boldsymbol{\nu},\boldsymbol{\nu})\\
\end{split}
\end{equation}

\section{Derivatives of the Hirsch vector field as function of potential}
In the following section we derive explicit formulation of Lie derivatives of  one-parameter diffeomorfism vector fiel $\boldsymbol{\xi}$ for a potential $V$ in "critical points-free" region of configuration space $(\mathcal{X},g_{\mathbb{R}^N}$ endowed with a riemmanian metric. Let $(q_1,....,q_N)$ be a set of coordinates in configuration space; in what follows we shall refer to  $\partial_i$ as the partial derivatives
respect to coordinate $q_i$ and (with an abuse of notation respect to the main part of this manuscript)
$(\nabla V)_i=(\mathbf{grad}_{\mathbb{R}^N}V)_i=\partial_i V$ and the Hessian $(\Hess V)_{ij}=\partial^2_{ij}V$.\\
With these chioces the divergence of Hirsch vector field $\zeta=\mathrm{div}_{\mathbb{R}^N}\boldsymbol{\xi}$ reads:
\begin{equation}
\mathrm{div}_{\mathbb{R}^N}\boldsymbol{\xi}=\dfrac{\Delta V}{\|\nabla V\|^2}-2\dfrac{\nabla V \cdot(\Hess V\nabla V)}{\|\nabla V\|^4}
\end{equation}
where $\Delta(\cdot)=\sum_{i}^N \partial^{i}\partial_{i}(\cdot)$ is the Laplacian operator in the Euclidean configuration space and $\|\boldsymbol X\|^2=g_{\mathbb{R}^N}(\boldsymbol X,\boldsymbol X)$ is the Euclidean norm. Consequentely we calculate explicitely higher order Lie derivative of $\zeta$ respect to $\boldsymbol{\xi}$
as averages, correlations, and other cumulants of this quantities appears in calculation of microcanonical entropy density. As the Lie derivative operator along the flux generated by vector field $\boldsymbol{\xi}$ is 
\begin{equation}
\mathcal{L}_{\boldsymbol{\xi}}(\cdot)=(\boldsymbol{\xi}\cdot\nabla)(\cdot)=\sum_{i=1}^{N}\dfrac{\partial^{i}V}{\|\nabla V\|^2}\partial_i(\cdot)=\dfrac{\partial^i V}{\|\nabla V\|^2}
\end{equation}
This yields at the first order:
\begin{equation}
\begin{split}
\mathcal{L}_{\boldsymbol{\xi}}(\zeta)=&\dfrac{\nabla V\cdot\nabla(\Delta V)}{\|\nabla V\|^4}-2\dfrac{(\nabla V \cdot\Hess(V)\nabla V)\Delta V+2\|\Hess V\nabla V\|^2+\mathrm{D}^3V(\nabla V,\nabla V,\nabla V)}{\|\nabla V\|^6}+\\
&+8\dfrac{(\nabla V \cdot\Hess V\nabla V)^2}{\|\nabla V\|^8}
\end{split}
\end{equation}
at the second order:
\begin{equation}
\begin{split}
\mathcal{L}_{\boldsymbol{\xi}}^{(ii)}(\zeta)=& \dfrac{\nabla(\Delta V)\cdot (\Hess V \nabla V)+\nabla V \cdot(\Hess(\Delta V)\nabla V)}{\|\nabla V\|^6}+\\
&-2\Biggr[\dfrac{\Delta V\mathrm{D}^3 V(\nabla V,\nabla V,\nabla V)+2\Delta V\|\Hess V\nabla V\|^2}{\|\nabla V\|^8}+\\
&+\dfrac{4(\Hess V\nabla V)\cdot(\Hess V\Hess V\nabla V)}{\|\nabla V\|^8}+\\
&+\dfrac{7\mathrm{D}^3V(\Hess V\nabla V,\nabla V,\nabla V)+\mathrm{D}^4V(\nabla V,\nabla V,\nabla V,\nabla V)}{{\|\nabla V\|^8}¨+3 (\nabla V\cdot \Hess V\nabla V)(\nabla V\cdot \nabla(\Delta V))}{\|\nabla V\|^8}\Biggr]+\\
&+\dfrac{28(\nabla V\Hess V\nabla V)\left[2\|\Hess V \nabla V\|^2+\mathrm{D}^3V(\nabla V,\nabla V,\nabla V)\right]+12(\nabla V\Hess V\nabla V)^2\Delta V}{\|\nabla V\|^{10}}+\\
&-64\dfrac{(\nabla V\Hess V\nabla V)^3}{\|\nabla V\|^{12}}
\end{split}
\end{equation}
at the third order:
\begin{equation}
\begin{split}
&\mathcal{L}_{\boldsymbol{\xi}}^{(iii)}(\zeta)=\dfrac{3 \nabla V \cdot\Hess(\Delta V)\Hess V\nabla V+\mathrm{D}^3\Delta V(\nabla V,\nabla V,\nabla V)+\mathrm{D}^{3}V(\nabla V,\nabla V,\nabla(\Delta V))}{\|\nabla V\|^8}+\\
&+\dfrac{\nabla (\Delta V)\cdot \Hess V\Hess V \nabla V}{\|\nabla V\|^8}-2\Biggr[\dfrac{4 \mathrm{D}^{3}V(\nabla V,\nabla V,\nabla V)(\nabla V \cdot \nabla(\Delta V))}{\|\nabla V\|^{10}}+\\
&+\dfrac{
7 \mathrm{D}^{4}V(\nabla V,\nabla V,\nabla V,\Hess V\nabla V)}{\|\nabla V\|^{10}}+\\
&+\dfrac{15 \mathrm{D}^{3}V(\nabla V,\nabla V,\Hess V \Hess V \nabla V)+7\|\mathrm{D}^{3}V(\nabla V,\nabla V)\|^2+18 \mathrm{D}^{3}V(\Hess V\nabla V,\Hess V \nabla V,\nabla V)}{\|\nabla V\|^{10}}+\\
&+\dfrac{4\mathrm{D}V(\nabla V,\nabla V, \nabla V,\Hess V \nabla V)+\mathrm{D}^{5}V(\nabla V, \nabla V, \nabla V,\nabla V, \nabla V)+8(\nabla V \cdot \nabla (\Delta V))\|\Hess V \nabla V\|^2}{\|\nabla V\|^{10}}+\\
&+\dfrac{8\|\Hess V \Hess V\nabla V\|^2+7\mathrm{D}^3 V(\nabla V,\nabla V,\Hess V \nabla V)\Delta V}{\|\nabla V\|^{10}}+\\
&\dfrac{\Delta V\mathrm{D}^{4}V(\nabla V,\nabla V,\nabla V,\nabla V)+4\Delta V (\Hess V \nabla V)\cdot \Hess V \Hess V \nabla V}{\|\nabla V\|^{10}}+\\
&+\dfrac{6(\Hess V \cdot \Hess V \nabla V)(\nabla V\cdot \Hess (\Delta V)\nabla V)+6(\Hess V \cdot \Hess V \nabla V)(\nabla(\Delta V)\cdot \Hess V \nabla V)}{\|\nabla V\|^{10}}\Biggr]+\\
&+4\Biggr[\dfrac{7\left(\mathrm{D}^3 V(\nabla V,\nabla V, \nabla V)\right)^2+28 \mathrm{D}^{3}V(\nabla V, \nabla V, \nabla V) \|\Hess V\nabla V\|^2}{\|\nabla V\|^{12}}+\\
&+\dfrac{10 \Delta V \mathrm{D}^3 V(\nabla V, \nabla V,\nabla V)(\nabla V \cdot \Hess V \nabla V)}{\|\nabla V\|^{12}}+\\
&+\dfrac{28 \|\Hess V \nabla V\|^4+20 \Delta V\|\Hess V \nabla V\|^2(\nabla V\cdot \Hess V \nabla V)}{\|\nabla V\|^{12}}+\\
&+\dfrac{(\nabla V\cdot \Hess V \nabla V)[77 \mathrm{D}^3 V(\nabla V,\nabla V,\Hess V \nabla V)}{\|\nabla V\|^{12}}+\\
&+\dfrac{11 \mathrm{D}^4 V(\nabla V,\nabla V,\nabla V,\nabla V)+44 (\Hess V \nabla V)\cdot(\Hess V \Hess V \nabla V)}{\|\nabla V\|^{12}}+\\
&+\dfrac{15 (\Hess V \cdot \Hess V \nabla V)(\nabla V\cdot \nabla(\Delta V))]}{\|\nabla V\|^{12}}\Biggr]+\\
&-8\Biggr[\dfrac{59\mathrm{D}^{3}V(\nabla V,\nabla V,\nabla V)(\nabla V \cdot \Hess V \nabla V)^2}{\|\nabla V\|^{14}}+\\
&+\dfrac{(\nabla V  \cdot \Hess V \nabla V)^2 [118 \|\Hess V \nabla V\|^2+15 \Delta V (\nabla V  \cdot \Hess V \nabla V)]}{\|\nabla V\|^{14}}\Biggr]+768\dfrac{(\nabla V  \cdot \Hess V \nabla V)^4}{\|\nabla V\|^{16}}
\end{split}
\end{equation}

\bibliography{sections/PhD_Biblio.bib}

\end{document}